\documentclass[usenatbib]{mn2e}
\usepackage{epsfig}
\usepackage{subfigure}
\usepackage{amsmath,amssymb}
\usepackage{natbib}
\usepackage{color}
\usepackage{aas_macros}  

\begin{document}

\title[Discovery of two lead-rich hot subdwarfs]
{Discovery of extremely lead-rich subdwarfs: does heavy metal signal the formation of subdwarf B stars?}

\author[Naslim, N., et al.]
{
Naslim~N.$^{1,2}$\thanks{E-mail: naslimn@asiaa.sinica.edu.tw}, 
C.~S.~Jeffery$^1$\thanks{E-mail: csj@arm.ac.uk},
A.~Hibbert$^{1,3}$, and
N.~T.~Behara$^{1,4}$\\
$^1$Armagh Observatory, College Hill, Armagh BT61\,9DG\\
$^2$Institute of Astronomy and Astrophysics, Academica Sinica P.O Box 23-141, Taipei 10617, Taiwan R.O.C\\
$^3$School of Mathematics, Queens University Belfast, Belfast BT7 1NN\\
$^4$Institut d'Astronomie et d'Astrophysique, Universit\'e Libre de Bruxelles, B-1050 Bruxelles, Belgium
}

\date{Accepted .....
      Received ..... ;
      in original form .....}

\pagerange{\pageref{firstpage}--\pageref{lastpage}}
\pubyear{2013}
\label{firstpage}
\maketitle

\begin{abstract}
Hot subdwarfs represent a group of low-mass helium-burning stars
formed through binary-star interactions and include some of the most
chemically-peculiar stars in the Galaxy.  Stellar evolution theory
suggests that they should have helium-rich atmospheres but, because
radiation causes hydrogen to diffuse upwards, a majority are extremely
helium poor.  Questions posed include: when does the atmosphere become
chemically stratified and at what rate?

The existence of several helium-rich subdwarfs suggests further
questions; are there distinct subgroups of hot subdwarf, or do hot
subdwarfs change their surface composition in the course of evolution?
Recent analyses have revealed remarkable surface chemistries amongst
the helium-rich subgroup.  In this paper, we analyse high-resolution
spectra of nine intermediate helium-rich hot subdwarfs. We report the
discovery that two stars, HE\,2359--2844 and HE\,1256--2738, show an
atmospheric abundance of lead which is nearly ten thousand times that
seen in the Sun. This is measured from optical Pb{\sc iv} absorption
lines never previously seen in any star.  The lead abundance is ten to
100 times that measured in normal hot subdwarf atmospheres from
ultraviolet spectroscopy.  HE\,2359--2844 also shows zirconium and
yttrium abundances similar to those in the zirconium star LS
IV$-14^{\circ}116$.  The new discoveries are interpreted in terms of
heavily stratified atmospheres and the general picture of a surface
chemistry in transition from a new-born helium-rich subdwarf to a
normal helium-poor subdwarf.
\end{abstract}

\begin{keywords}
star: chemically peculiar (helium)
stars: evolution, 
stars: abundances,
stars: horizontal-branch,
stars: subdwarf,
atomic data
\end{keywords}

\section{Introduction}
The formation of hot subdwarf B stars remains a puzzle; 
they are observed as single stars, and as both close and wide binaries. 
They are widely regarded to be core-helium burners; the majority have 
hydrogen-rich atmospheres, but this is only a thin veneer, since  they behave
as helium main-sequence or extended horizontal-branch stars of approximately half 
a solar mass \citep{heber09}. The puzzle is that the majority are believed to be red-giant cores, 
stripped of their hydrogen envelopes, so at best their outer layers should be enriched in helium. 
Whilst the majority have
helium-poor surfaces (helium number fraction $n_{\rm He}<1\%$), a minority
show  helium-rich surfaces with a wide 
range of nitrogen and carbon abundances. It appears that in the
``normal'' (helium-poor) sdB stars,  radiative levitation and gravitational
settling  cause helium to sink below the
hydrogen-rich surface \citep{heber86}, deplete other light
elements, and enhance many heavy elements in the
photosphere \citep{otoole06}.

It has been found that almost 10$\%$ of the total subdwarf population
comprises stars with helium-rich atmospheres \citep{green86,ahmad06,nemeth12}.
These are sometimes referred to as He-sdB and He-sdO stars, depending on the ratios
of certain He\,{\sc i}  and He\,{\sc ii}  lines \citep{moehler90,ahmad04a,drilling13} 
or more generally as helium-rich hot subdwarfs (He-sd's). 
A small number of these show a  surface helium 
abundance in the range $n_{\rm He} \approx 5 - 80\%$. 
\citet{naslim12} suggested a terminology based on helium content.  
He-sd's with  $n_{\rm He} >80\%$ were described as extremely helium-rich, 
whilst those having $5\% < n_{\rm He} <80\%$ were described as
intermediate helium-rich. Since the numbers in both groups are small, 
the question naturally arises whether these are truly distinct classes, 
or simply a convenient description of an undersampled continuum. 
The corollary is whether the classes represent different 
stages in the evolution of similar objects, thus representing a gradual
change in photospheric composition due to slow chemical separation, 
or represent objects with quite distinct origins. 

The formation of the extreme-helium subdwarfs appears to be well
explained by the merger of two helium white dwarfs
\citep{zhang12a}. However,  it is harder to understand 
the intermediate-helium subdwarfs; to date, few have been analyzed and 
those that have been are  diverse. 
The latter include the prototype JL\,87 \citep{ahmad07}, the
zirconium star LS\,IV$-14^{\circ}116$ \citep{naslim11}, the short
period binary CPD$-20^{\circ}1123$ \citep{naslim12}, 
and also UVO\,0512--08 and PG\,0909+276 \citep{edelmann03th}.
They appear to occupy a region of effective-temperature -- helium-abundance
space which is almost unpopulated in the  \citet{nemeth12} survey (Fig. 6). 
The question posed by these stars
is whether they are related either to normal sdB stars or to 
extreme He-subdwarfs or to both. The question can be addressed
by, {\it inter alia}, establishing whether  the variation in chemical 
and/or binary properties across the three groups is discrete or continuous. 

Although rare, a few apparently intermediate-helium subdwarfs were 
identified and partially analysed in the course of the 
ESO Supernova Ia Progenitor surveY  (SPY) 
\citep{napiwotzki01,lisker04,stroeer07,hirsch09}. 
Our present objective was to analyse nine SPY He-sd's in greater detail and
attempt to answer the question just posed.

\section{Observations}

The ESO SPY \citep{napiwotzki01} obtained VLT/UVES spectra for 
76 sdB/sdOB and 58 sdO stars \citep{lisker04} which had been identified as
white dwarf candidates mostly from the Hamburg ESO survey \citep{christlieb01}. 
Reduced high-resolution spectra were obtained from the ESO UVES 
archive \citep{ballester00}  for 
HE\,0111--1526, 
HE\,1135--1134, 
HE\,1136--2504, 
HE\,1238--1745 
HE\,1256--2738, 
HE\,1258+0113,
HE\,1310--2733, 
HE\,2218--2026 
and 
HE\,2359--2844. 
These had been indentified as having 
$0.05 < n_{\rm He} < 0.90$  by \citet{stroeer07}. 
For each star at least two spectra were available with signal-to-nose ratios 
between 26 and 31 in the continuum. 
For abundance analysis we selected the wavelength range  3600 -- 5000\, \AA . 
The UVES spectra of all nine intermediate He-sds display strong lines of interesting ions. 

\noindent{\it Carbon.} HE\, 2359--2844, HE\,0111--1526,
HE\,2218--2026 and HE\,1256--2738 show strong C{\sc ii} and C{\sc iii}
lines. No carbon lines were identified in HE\,1310--2733,
HE\,1135--1134, HE\,1136--2504, HE\,1238--1745 and
HE\,1258+0113. 

\noindent{\it Zirconium.} \citet{naslim11} 
reported  zirconium and yttrium lines in LS\,IV$-14^{\circ}116$. 
The same Zr{\sc iv} and Y{\sc iii} lines along with two more Zr{\sc iv} lines at 3687 and 3764 \AA\  are
found in HE\, 2359--2844. 

\noindent{\it Lead.} The UVES spectrum of HE\, 2359--2844
shows a strong absorption line at 4049.8\AA\ and
weaker lines at 3962.5 and 4496.1\AA. These have been identified from the NIST atomic database 
to be due to Pb {\sc iv} \citep{nist12}. 
HE\,1256--2738 also shows the Pb {\sc iv} 4049.8\AA\ line and 
3962.5; 4496.1\AA\ is too weak to measure. 
To our knowledge, these lines have not been observed in any 
other astronomical object, although \citet{otoole04} identified Pb{\sc iv} lines in  
Space Telescope ultraviolet spectra of the sdB stars  Feige 48 and PG1219+534.

\begin{table*}
\caption{Atmospheric parameters}
\label{t_pars}
\begin{tabular}{@{}lllllll}
\hline
Star & $T_{\rm eff} (\rm K)$ & $\log g$ & $n_{\rm He}$ & $y$ &  $v \sin i$ & Source\\
     &                      &          &              &    & $({\rm km\,s^{-1}})$ & \\
\hline
HE\,0111--1526
& $38\,310\pm1200$   & $5.93\pm0.2$ & $0.80\pm0.12$ &  $4.0\pm0.6$  &     $3\pm1$         & SFIT     \\
& $39\,152$         & $6.31$        & 0.87 &       &    & 1 \\[1mm]
 HE\,1135--1134
& $38\,400\pm1500$   & $5.65\pm0.1$ & $0.36\pm0.15$ & $0.56\pm0.23$ &   $2\pm1 $         & SFIT    \\
& $40\,079$         & $5.68$        & 0.35 &       &   & 1 \\
HE\,1136--2504
& $40\,320\pm500$   & $5.65\pm0.05$ & $0.49\pm0.19$ & $0.96\pm0.37$ &     $3\pm2$        & SFIT    \\
& $41\,381$         & $5.84$        & 0.40 &      &    &  1\\
HE\,1238--1745
& $37\,230\pm800$   & $5.57\pm0.2$ & $0.28\pm0.05$ & $0.39\pm0.07$ &   $8\pm2$          & SFIT    \\
& $38\,219$         & $5.64$        & 0.22 &      &    & 1 \\
HE\,1256--2738 
& $39\,500\pm1000  $ & $5.66\pm0.1$ & $0.49\pm0.19$ & $0.96\pm0.37$ &      $4\pm1$       & SFIT \\
& $40\,290  $       & $5.68$        & 0.55      &     &        &1 \\[1mm]
HE\,1258+0113
& $38\,780\pm500$   & $5.56\pm0.12$ & $0.25\pm0.1$ & $0.33\pm0.13$ &    $5\pm1 $        & SFIT    \\
& $39\,359$         & $5.64$        & 0.23 &       &   & 1 \\[1mm]
HE\,1310--2733
& $38\,400\pm1100$ & $5.48\pm0.15$ & $0.44\pm0.15$ & $0.79\pm0.27$ &    $6\pm2$         & SFIT   \\
& $40\,000$         & $5.63$        & 0.41 &      &    & 1 \\[1mm]
 HE\,2218--2026
& $37\,280\pm1500$   & $5.8\pm0.1$  & $0.30\pm0.1$ & $0.43\pm0.14$ & $10\pm3$   & SFIT \\
& $38\,330  $       & $5.87$        &  0.31     &      &       & 1 \\[1mm]
 HE\,2359--2844
& $37\,050\pm1000$   & $5.57\pm0.15$  & $0.43\pm0.18$ & $0.75\pm0.32$ &  $5\pm2$        & SFIT \\
& $38\,325$          & $5.65$        & 0.42 &        &     & 1  \\
\hline
\end{tabular}\\
\parbox{80mm}{
Reference: 
1. \citet{stroeer07}
}
\end{table*}

\begin{figure}
\centering \epsfig{file=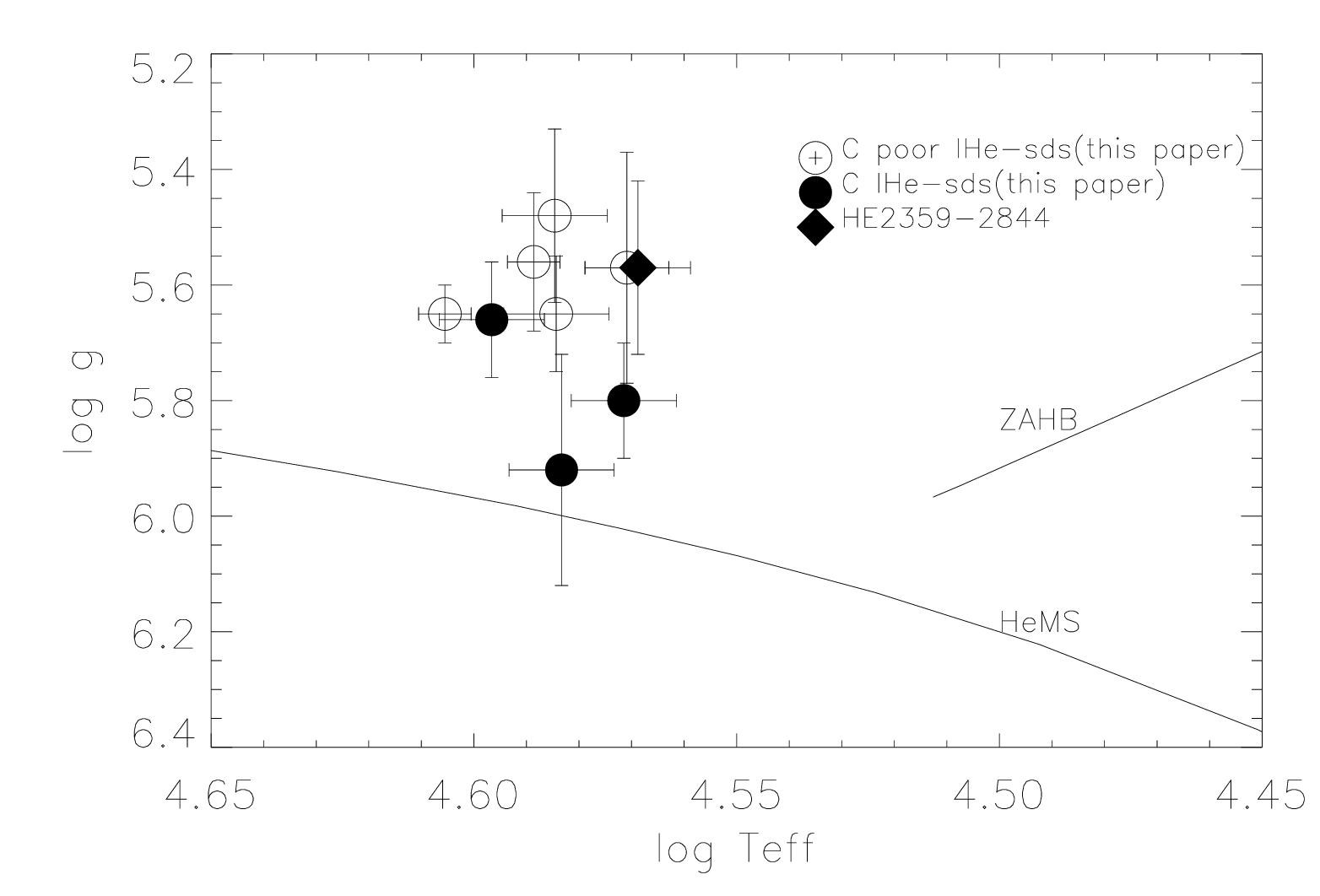, angle=0,
  width=8cm,height=6cm}
  \caption{$T_{\rm eff}$ and $\log g$ for nine intermediate
    He-subdwarfs measured for this paper. Symbols distinguish
    carbon-rich and carbon-poor stars, as well as the zirconium star
    HE\,2359--2844. Approximate locations for the helium main sequence
    and the zero-age horizontal branch are also shown. }
  \label{teff-log1}  
\end{figure}

\begin{figure}
\centering \epsfig{file=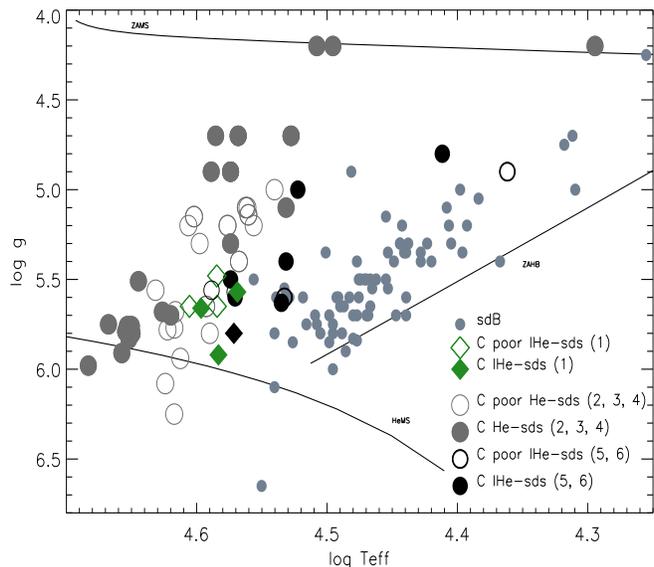, angle=0,
  width=8.5cm, height=7.3cm}
  \caption{Comparison of C-rich He-sd's and C-poor He-sd's in a
    $T_{\rm eff}$-$\log g$ plane. The hot subdwarfs shown in this
    figure include 1. Intermediate He-sd's  discussed in this
    paper, 2. He-sd's from\citet{naslim10}, 3. Intermediate He-sd's
    from \citet{naslim11}, 4. He-sdO's from \citet{stroeer07}, 5. He-sd's
    from \citet{ahmad03}, and normal sdBs from \citet{edelmann03}.}
  \label{teff-log2}  
\end{figure}

\section{Atmospheric parameters}

We measured effective temperature $T_{\rm eff}$, surface gravity $\log
g$, and helium abundance $n_{\rm He}$ by fitting He {\sc i} and Balmer
lines using the $\chi^{2}$-minimization package SFIT
\citep{jeffery01}. The observed spectra were matched to a grid of
      {\it fully} line-blanketed models computed in
      local-thermodynamic and hydrostatic equilibrium using
      opacity-sampling and a linelist of some $10^6$ transitions
      \citep{behara06}. The grid covers a wide range in $T_{\rm eff}$, $g$,
and $n_{\rm He}$ for a number of distributions of elements heavier than helium, including solar,
1/10 solar and other custom-designed mixtures
\citep{behara06}.
 For this analysis we sought solutions in the range
$34\,000 < T_{\rm eff}/{\rm K} < 50\,000$, $4.5 < \log g (cgs) < 6.0$ and $0.10 < n_{\rm He} < 0.90$
over the wavelength interval $3600 - 5000$\AA.
A microturbulent velocity of $v_{\rm t}=10\,{\rm km\,s^{-1}}$ was adopted, as determined for other
      He-sdBs by \citet{naslim10}, where other details of the fitting
      procedure can also be found. 
The SFIT parameters $T_{\rm eff}$, $\log g$, $n_{\rm He}$ and 
$v \sin i$ (an upper limit to the projected rotational velocity) for each star are shown in
      Table~\ref{t_pars}, and also in Fig.~\ref{teff-log1}. 
        Errors given in Table~\ref{t_pars} are formal errors from the 
        $\chi^2$ solution. $y \equiv n_{\rm He}/n_{\rm H}$ is also included.  

 Since the measured values of $T_{\rm eff}$ lie close to $\approx
 40\,000$K, the increasing importance of departures from local
 thermodynamic equilibrium (LTE) needs to be considered and, in any
 case, our results should be compared with those obtained by
 \citet{stroeer07}.  The latter used a grid of partially
 line-blanketed non-LTE model atmospheres calculated using the code
 PRO2 \citep{werner86,werner03}. It remains to be seen whether the
 incompleteness of line-blanketing is more or less significant than
 departures from LTE for the stars considered here.  Our $T_{\rm eff}$
 are systematically cooler by $\approx 1200$\,K and our gravities are
 systematically weaker by $\approx 0.1$ dex than those of
 \citet{stroeer07}, but the helium abundances are in almost complete
 agreement. This gives some confidence in the abundances derived for
 other species using our models. The systematic increase in the
 logarithmic abundance of Pb{\sc iv} due to an increase in $T_{\rm
   eff}$ of 2\,000\,K is +0.12 for HE2359--2844 and --0.12 for
 HE1256--2738.

\begin{table*}
\centering
\caption{Elemental abundances in the form  $\epsilon_i = \log n_i + c$ (see text). The numbers of lines used in each measurement are shown in parentheses.}
\label{t_abunds}
\setlength{\tabcolsep}{3pt}
\begin{tabular}{@{\extracolsep{0pt}}llllllllll}
\hline
Star & H & He & C & N & O &Ne & Mg & Si & S    \\
\hline
HE\,0111--1526 &   10.88 & 11.48 & $8.49\pm0.23$ & $8.42\pm0.38$ & $<6.9$ &$7.51\pm0.23$ & $6.98\pm0.10$ & $6.79\pm0.31$ & $6.94\pm0.11$\\
               &         &       &     (18)          &      (20)         &               &    (6)    &    (1)        &      (7)       &    (2)     \\
HE\,1135--1134 &   11.67 & 11.42 & $<6.5      $ & $8.19\pm0.46$ & $<7.4$ &$<7.0$  & $<6.5$  &  $<6.1$  &$<6.4$\\
               &         &       &               &     (8)          &               &        &            &             &         \\
HE\,1136--2504 &   11.45 & 11.44 & $ <7.1    $   & $8.22\pm0.33$ & $<7.5$  & $<7.1$   &  $<6.5$ & $<6.1$ &   $<6.6$  \\
               &         &       &               &       (12)        &               &        &            &             &         \\
HE\,1238--1745 &   11.72 & 11.31 &  $<6.4$  & $8.09\pm0.40$ &$<7.3$   & $<7.0$  &  $<6.5$ &   $5.97\pm0.12$ &    $6.97\pm0.12$ \\
               &         &       &               &     (11)          &               &        &            &    (2)         &    (2)     \\
HE\,1256--2738 &   11.45 & 11.44 & $8.90\pm0.54$ & $8.14\pm0.62$ & $8.08\pm0.1$ &$<7.1$ & $<6.5$ &  $6.19\pm0.10$ & $<6.5$\\[1mm]
               &         &       &      (17)         &    (10)           &    (3)           &        &            &    (2)         &         \\
HE\,1258+0113  &   11.74 & 11.26 &  $<6.9$ & $7.59\pm0.42$  & $<7.4$  &$<7.0$  &  $<6.5$  & $<6.0$   & $<6.4$     \\
               &         &       &               &   (5)            &               &        &            &             &         \\
HE\,1310--2733 &   11.49 & 11.39 & $<6.8$         & $8.29\pm0.28$   &$<7.3$   &$7.76\pm0.1$  &$7.76\pm0.09 $ & $6.80\pm0.08$ &$6.81\pm0.05$\\
               &         &       &               &      (18)         &               &  (5)      &    (1)        &    (6)         &   (2)      \\
HE\,2218--2026 &   11.60 & 11.22 & $8.81\pm0.83$ & $<7.2$  & $<7.5$  &$<7.1$ &  $<6.5$  &  $<6.1$   & $<6.5$\\
               &         &       &      (9)         &               &               &        &            &             &         \\
HE\,2359--2844 &   11.58 & 11.38 & $8.51\pm0.29$ & $8.00\pm0.57$ & $7.81\pm0.16$ &$<6.9$ & $7.6\pm0.1$   &$5.73\pm0.13$ & $<6.3$\\
               &         &       &    (15)           &     (9)          &     (5)          &        &            &     (2)        &         \\[1mm]
JL\,87$^{1}$       &    $11.62\pm0.07$ & $11.26\pm0.18$ & $8.83\pm0.04$ & $8.77\pm0.23$  & $8.6\pm0.23$ &$8.31\pm0.57$ & $7.36\pm0.33$&$7.22\pm0.27$     &$6.88\pm1.42$\\
LS\,IV$-14^{\circ}116^{2}$ & 11.83 & $11.23\pm0.05$ & $8.04\pm0.22$ & $8.02\pm0.2$  & $7.6\pm0.17$ & $<7.6$ & $6.85\pm0.1$ & $6.32\pm0.12$ &\\[1mm]
Sun$^{3}$      & 12.00     &[10.93]& 8.43 & 7.83  &8.69  & [7.93] &  7.60  & 7.51 & 7.12   \\
\hline
\end{tabular}\\
\parbox{160mm}{
References: 
1. \citet{ahmad07},
2. \citet{naslim11}, 
3. \citet{asplund09}; photospheric abundances except helium (helioseismic) and neon. 
}
\end{table*}

\begin{table*}
\centering
\caption{Abundances derived for Zr {\sc iv}, Y {\sc iii}, Pb {\sc iv} in HE\,2359--2844 and Pb {\sc iv} in HE\,1256--2738.}
\label{t_abunds2}
\setlength{\tabcolsep}{3pt}
\begin{tabular}{@{\extracolsep{0pt}}ccllrlrlll}
\hline
Ion &  &  & & \multicolumn{2}{c}{HE\,2359--2844} & \multicolumn{2}{c}{HE\,1256--2738} & LS\,IV$-14^{\circ}116$ &  Sun\\
$\lambda/{\rm \AA}$ & Configuration & $E_{i}/{\rm eV}$ & $\log
gf$ &$w_{\lambda}/{\rm m\AA}$ & $\epsilon_{i}$ & $w_{\lambda}/{\rm m\AA}$ & $\epsilon_{i}$ & $\epsilon_{i}\,^3$ & $\epsilon_{i}\,^6$ \\
\hline

Zr {\sc iv}        &     &       &               &            &     &       \\
3686.905&$6p ^{2}P_{3/2}-6d ^{2}D_{5/2}$   & 21.17$^1$&  +0.746$^3$&  66 & 6.36 & & &\\
3764.319&$6d ^{2}D_{5/2}-6f ^{2}F_{7/2}$    & 24.51&  +0.485$^3$  & 26 & 6.61   & & &\\
4198.265&$5d^{2}D_{5/2}-6p^{2}P_{3/2}^{0}$& 18.23&   +0.323$^4$ & 86 & 6.32   & &  &\\
4317.081&$5d^{2}D_{3/2}-6p^{2}P_{1/2}^{0}$& 18.18&   +0.069$^4$ & 64 & 6.60   &  & &\\
        &                               &     &           & mean   &$6.47$ &  & & 6.53 & $2.58$\\
       &                               &     &           &         &$\pm0.15$ &  & & $\pm0.24$ & $\pm0.04$\\
Y {\sc iii}        &     &       &                      &     &       \\
4039.602&$4f^2F_{7/2} - 5g^2G_{9/2}$     & 12.53$^2$& $+1.005^3\rceil$ & 42 & 6.60 &  & & \\
4039.602&$4f^2F_{7/2} - 5g^2G_{7/2}$     & 12.53& $-0.538^3\rfloor$& &   & & &\\
4040.112&$4f^2F_{5/2} - 5g^2G_{7/2}$     & 12.53& $+0.892^4$   &          35 &  6.63  & & &\\[1mm]
        &                               &     &  &  mean  &$6.61$  & & & 6.16 & $2.21$   \\
       &                               &     &           &     &$\pm0.15$ & &  & $\pm0.10$ &  $\pm0.04$\\
Pb {\sc iv}        &     &       &                      &     &     & &  \\
 3962.48           & $6d^2D_{3/2} -5d^9 6s6p[16^o]_{1/2}$    &   22.88$^5$ & $-0.047^5$  &  15 &  5.45 &42 & 6.56 \\
 4049.80           & $7s^2S_{1/2} - 5d^9 6s6p[16^o]_{1/2}$    &   22.94      & $-0.065$&    25  & 5.74  &68 & 6.23 \\
 4496.15           & $6d^2D_{5/2} - 5d^9 6s6p[15^o]_{3/2}$    &   23.16     &  $-0.237$&      15  & 5.73  & &  \\
       &         &     &  &      mean   &$5.64$  & &$6.39$ & & $1.75$ \\
     &         &     &  &                   &$\pm0.16$  & &$\pm0.23$ & & $\pm0.10$ \\
\hline
\end{tabular}\\
\parbox{150mm}{
References:
1. \citet{reader97},
2. \citet{epstein75},
3. \citet{naslim11}, 
4. this paper,\\
5. \citet{safranova04},
6. \citet{asplund09}.
}
\end{table*}

\begin{figure*}
 \includegraphics[trim=2cm 0cm 0cm 0cm, angle=-90,scale=0.3]{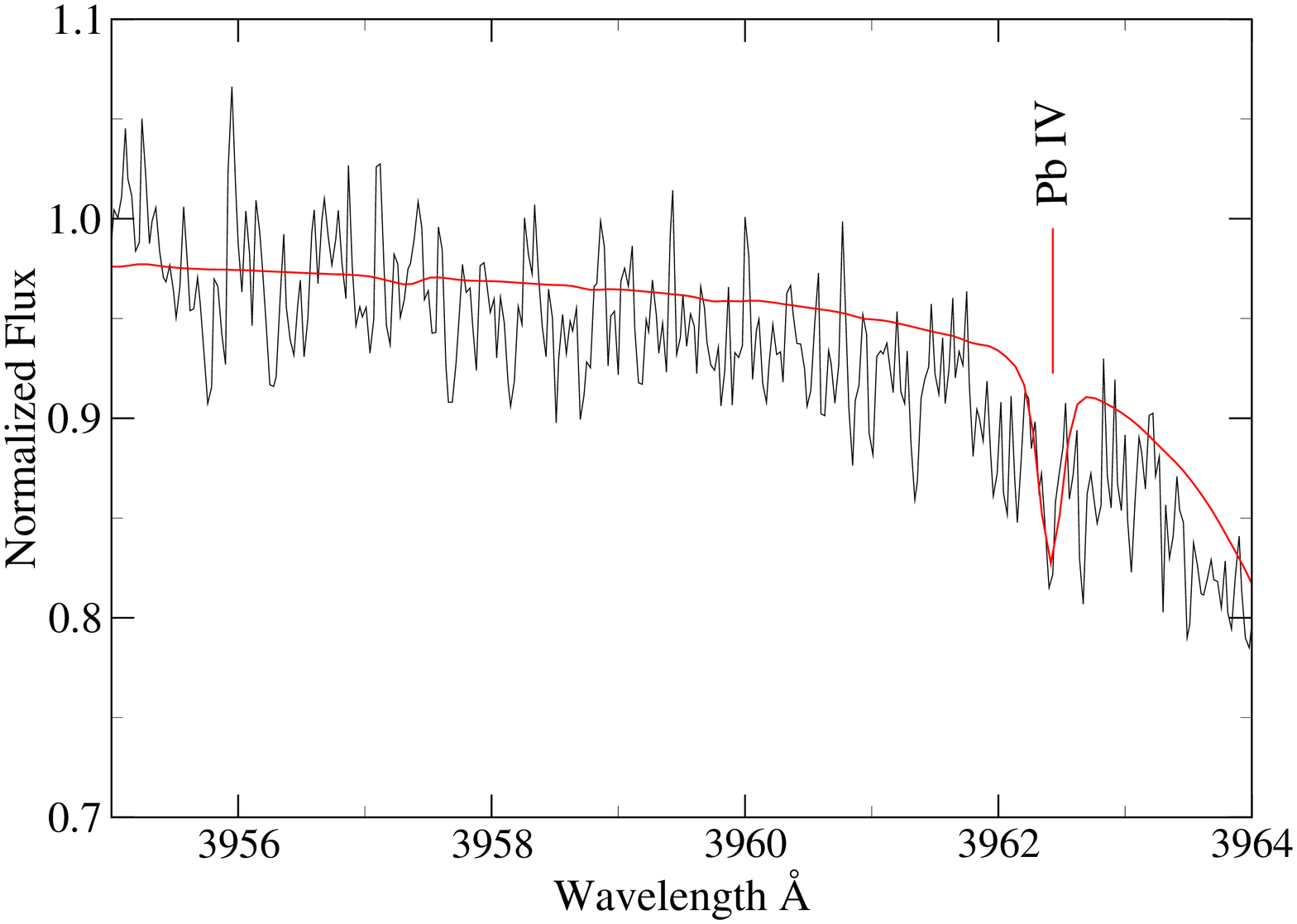}
 \includegraphics[trim=2cm 0cm 0cm 0cm, angle=-90,scale=0.3]{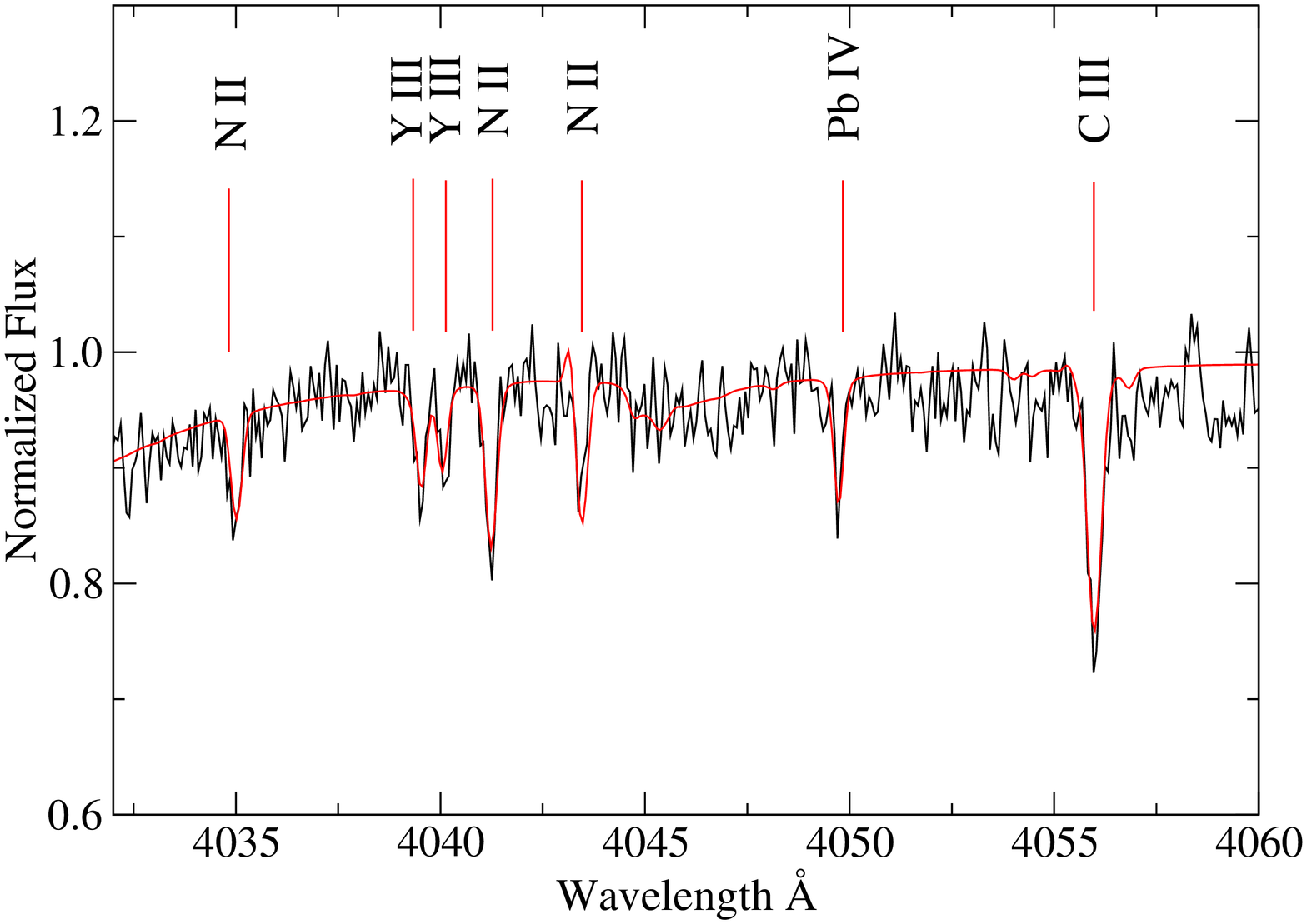}
\includegraphics[trim=2cm 0cm 0cm 0cm, clip=true, angle=-90,scale=0.3]{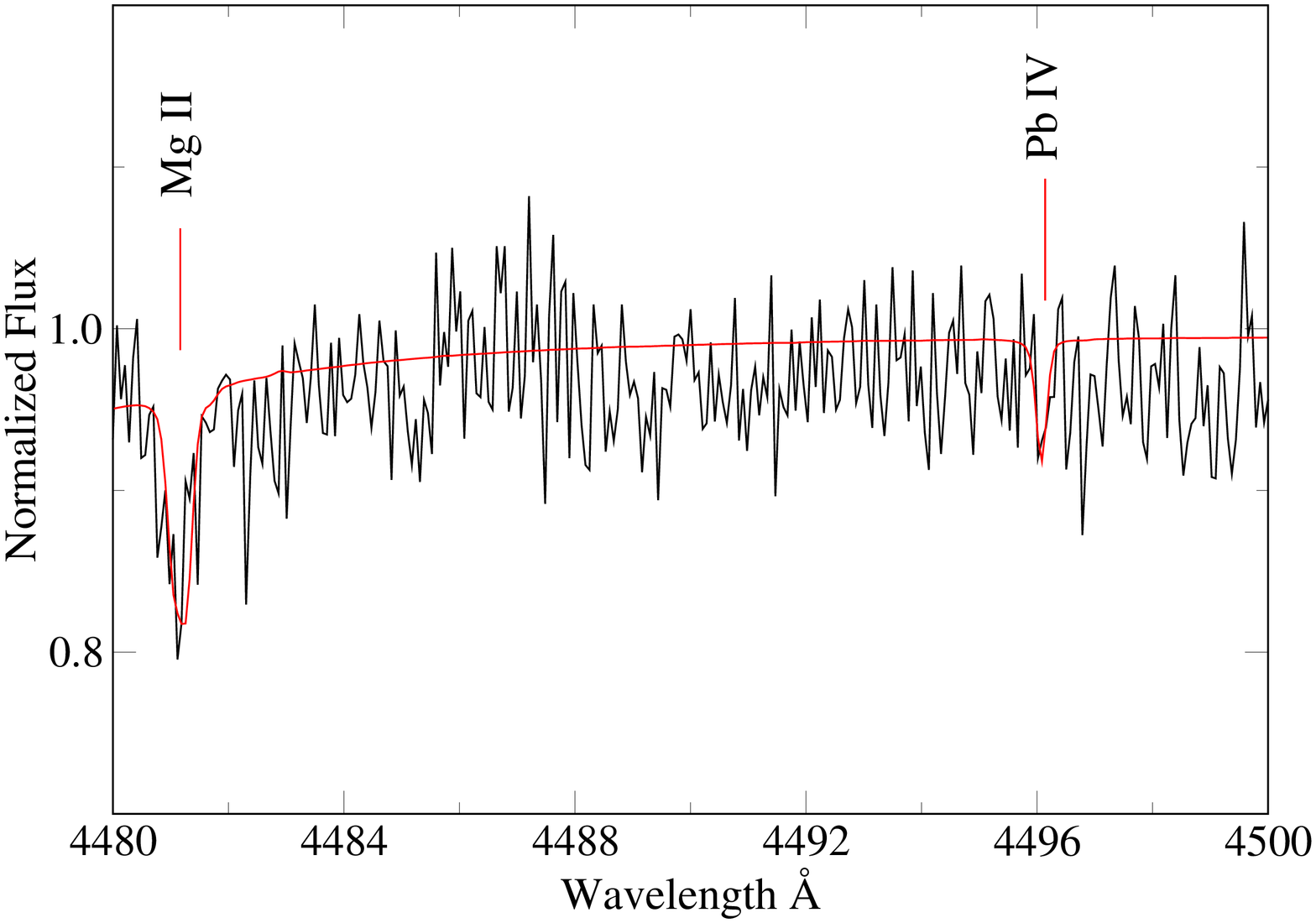}
\includegraphics[trim=2cm 0cm 0cm 0cm, clip=true, angle=-90,scale=0.3]{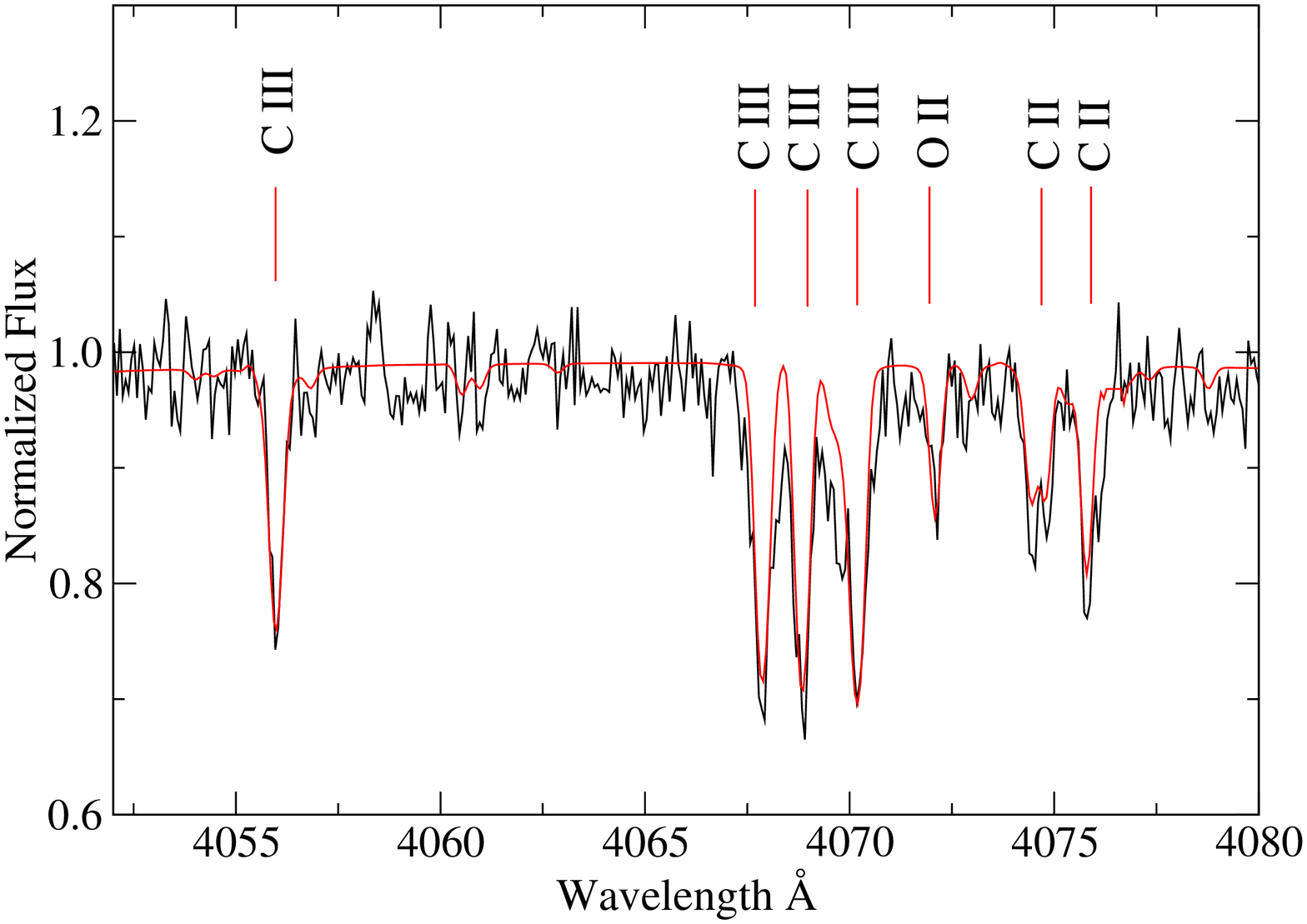}
\includegraphics[trim=2cm 0cm 0cm 0cm, clip=true, angle=-90,scale=0.3]{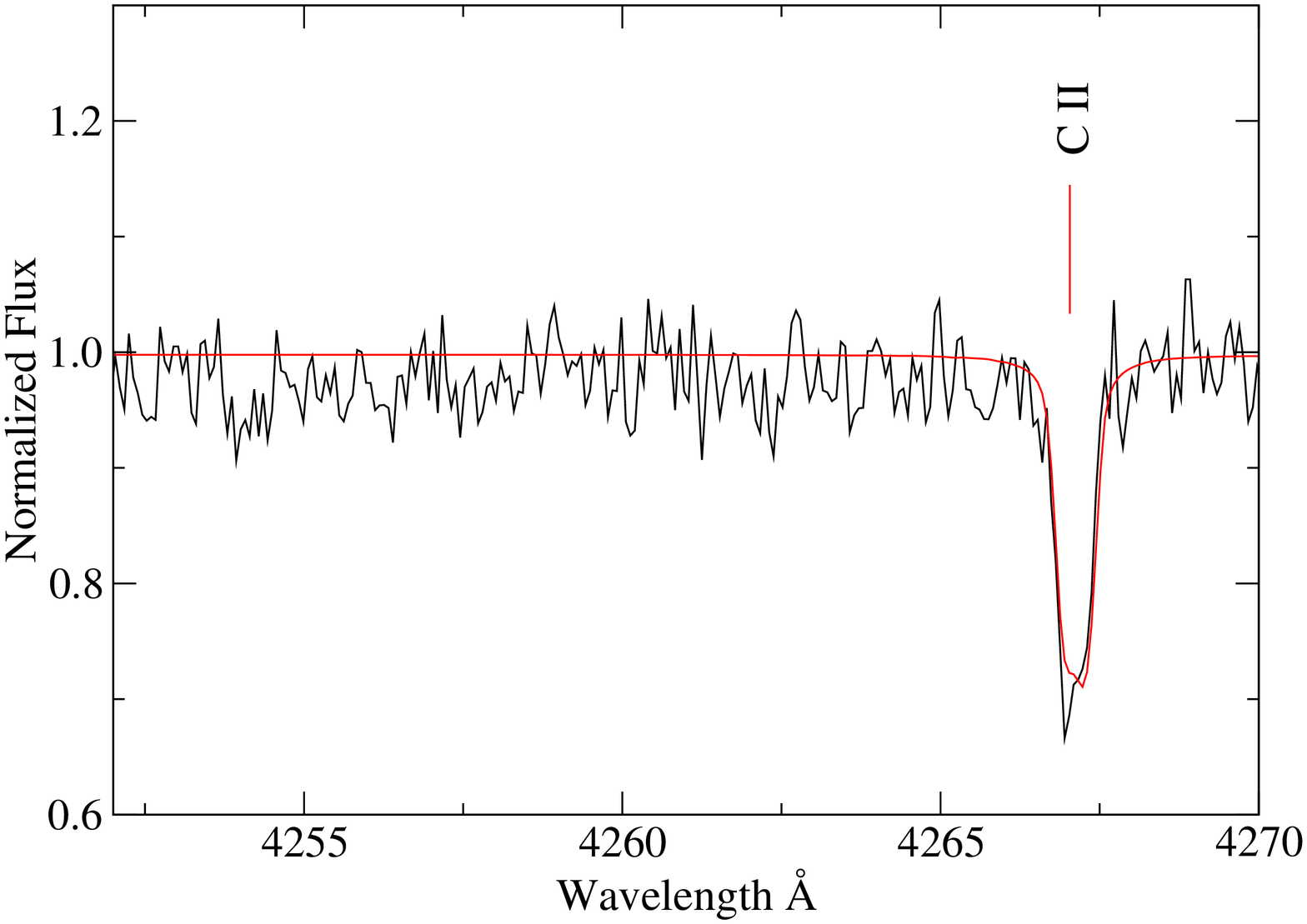}
\includegraphics[trim=2cm 0cm 0cm 0cm, clip=true, angle=-90,scale=0.3]{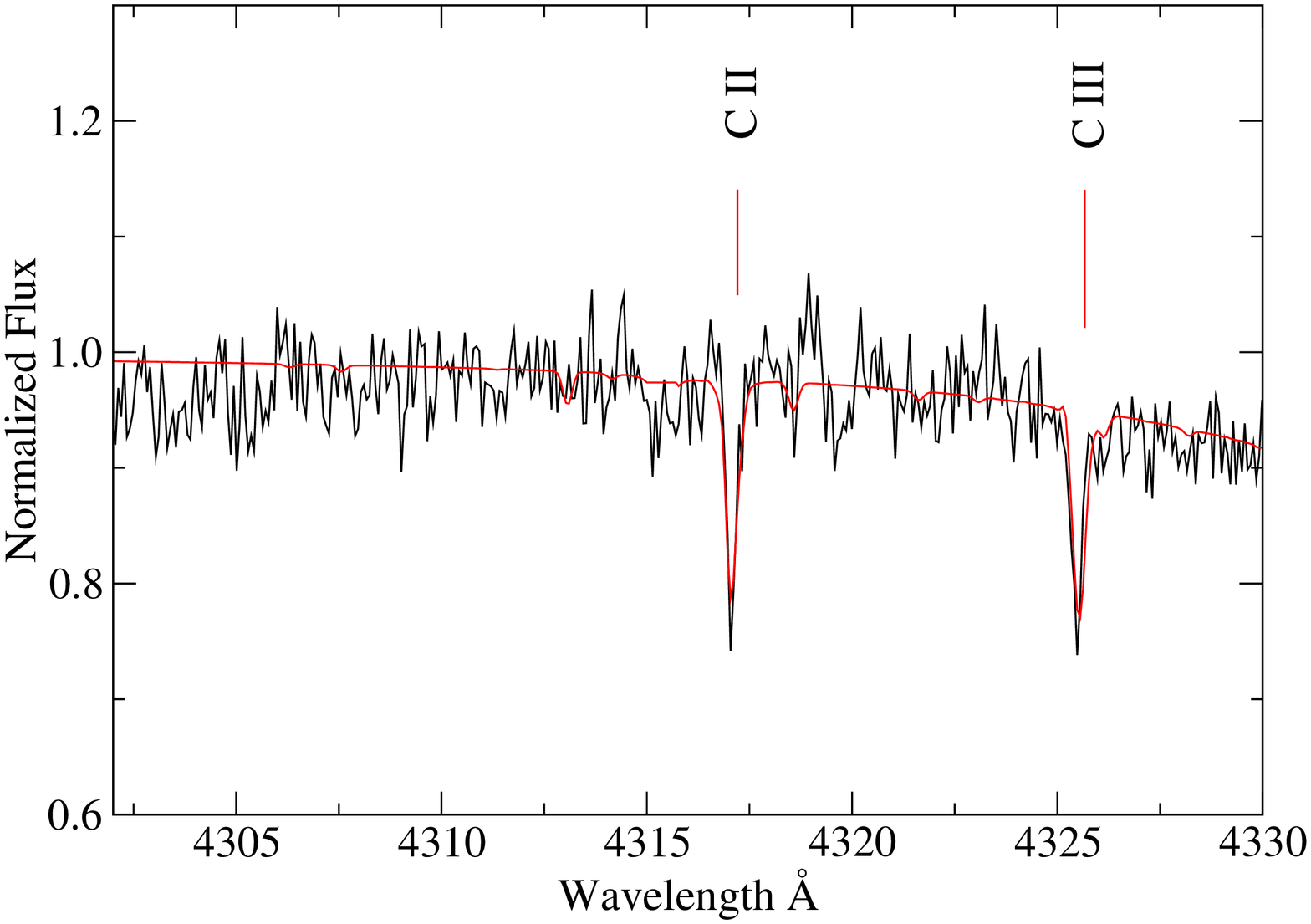}

\caption{Segments of the VLT UVES spectrum of HE\,2359--2844, 
together with the best-fit model, showing  C{\sc ii,iii}, N{\sc ii}, O{\sc ii}, Mg{\sc ii}, Y{\sc iii}
and Pb{\sc iv} lines. 
}
\label{f:2359a}
\end{figure*}

\begin{figure*}
\includegraphics[trim=1cm 0cm 0cm 0cm, clip=true, angle=-90,scale=0.3]{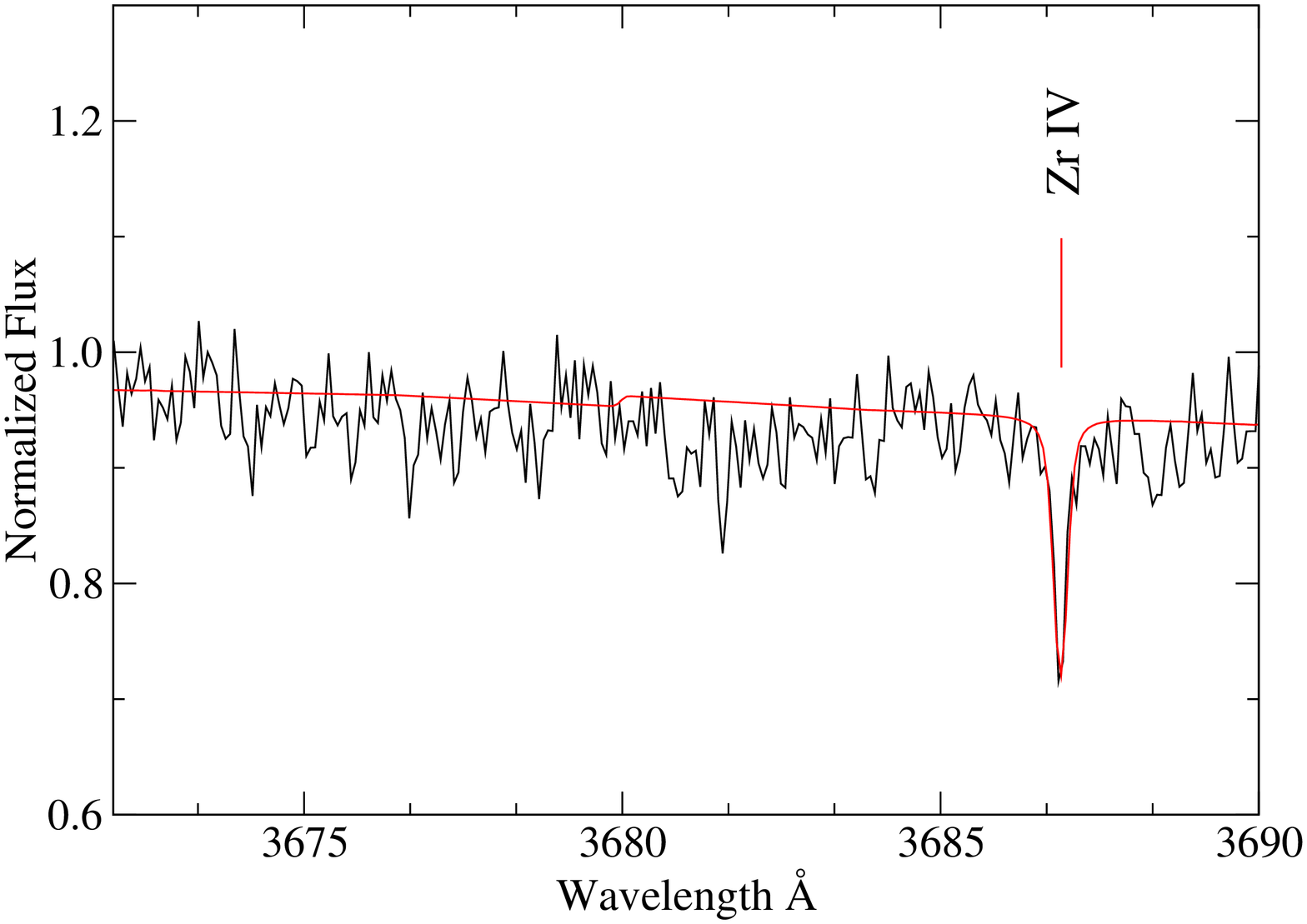}
\includegraphics[trim=1cm 0cm 0cm 0cm, clip=true, angle=-90,scale=0.3]{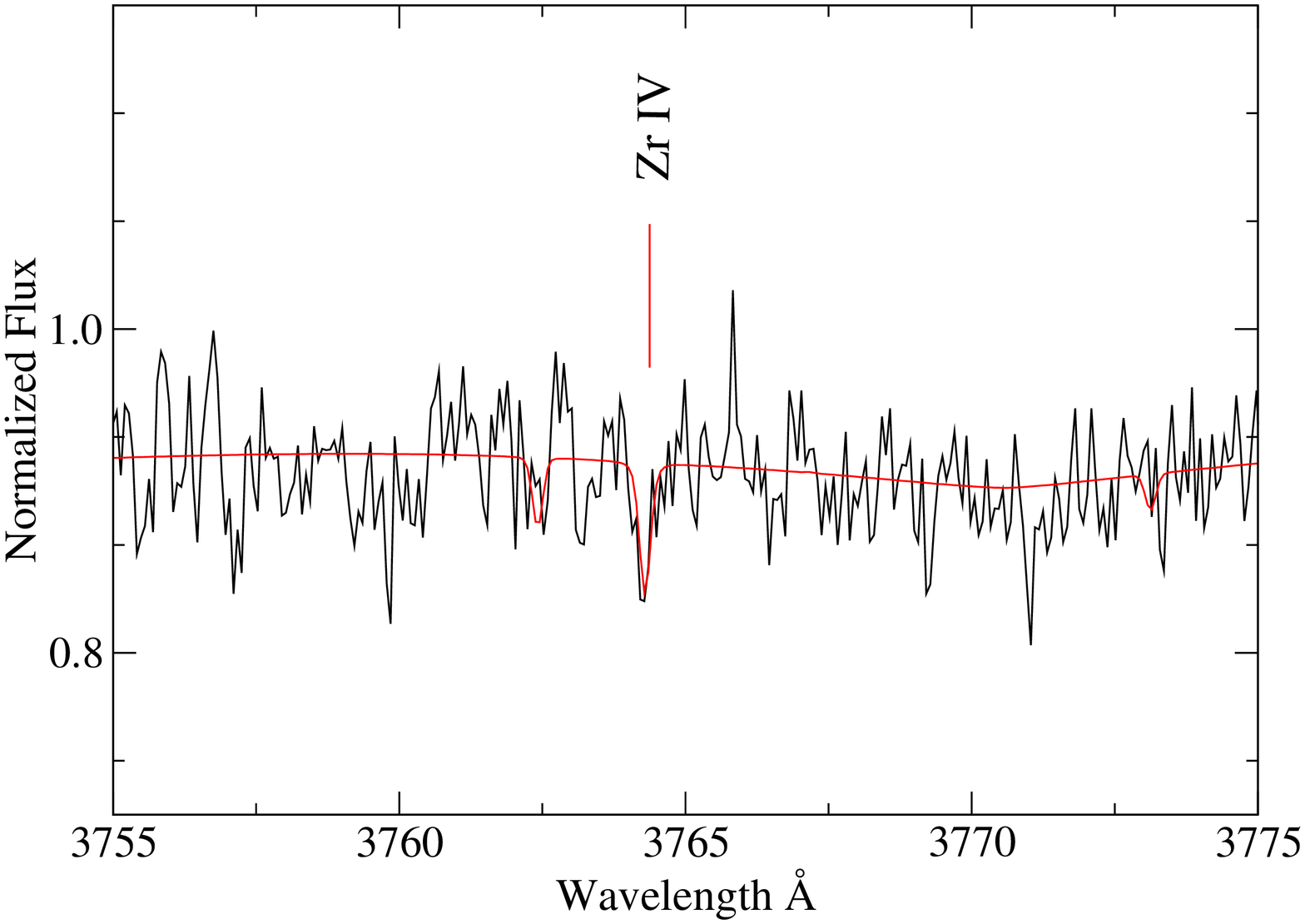}
\includegraphics[trim=1cm 0cm 0cm 0cm, clip=true, angle=-90,scale=0.3]{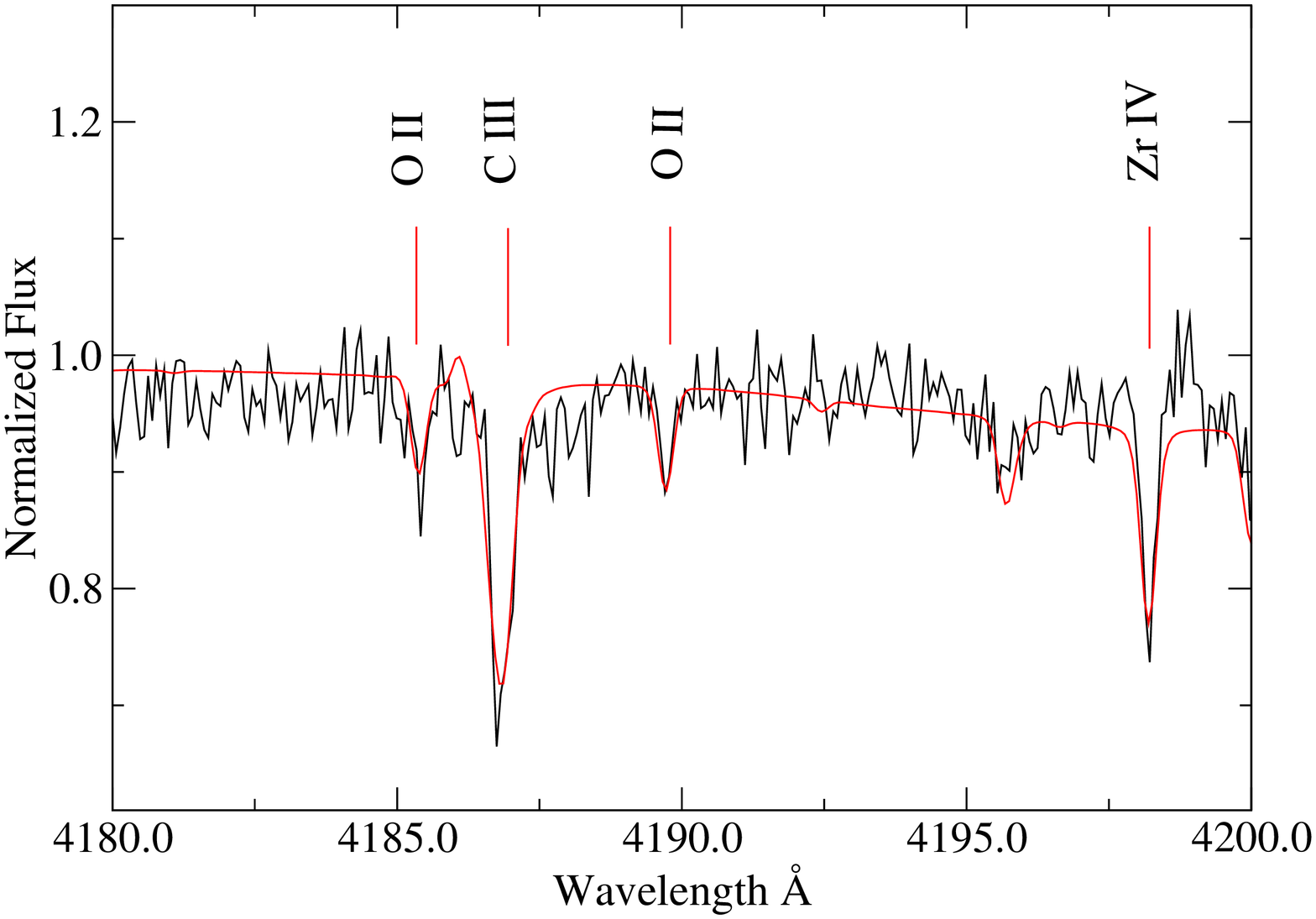}
\includegraphics[trim=1cm 0cm 0cm 0cm, clip=true, angle=-90,scale=0.3]{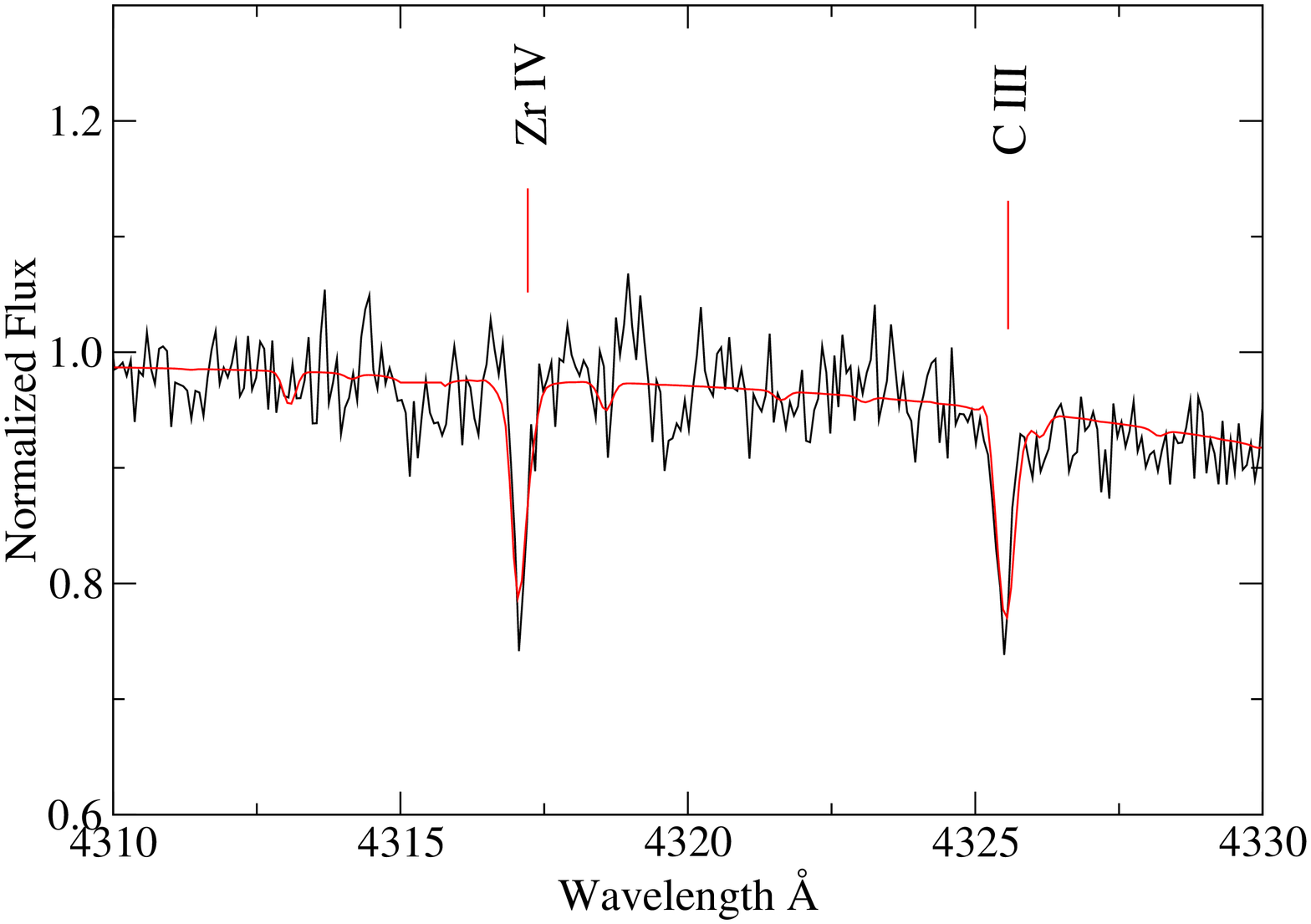}
\caption{Segments of the VLT UVES spectrum of HE\,2359--2844, 
together with the best-fit model, showing  C{\sc iii}, O{\sc ii} and Zr{\sc iv} lines. 
}
\label{f:2359b}
\end{figure*}

\begin{figure*}
\includegraphics[trim=2cm 0cm 0cm 0cm, clip=true,angle=-90,scale=0.3]{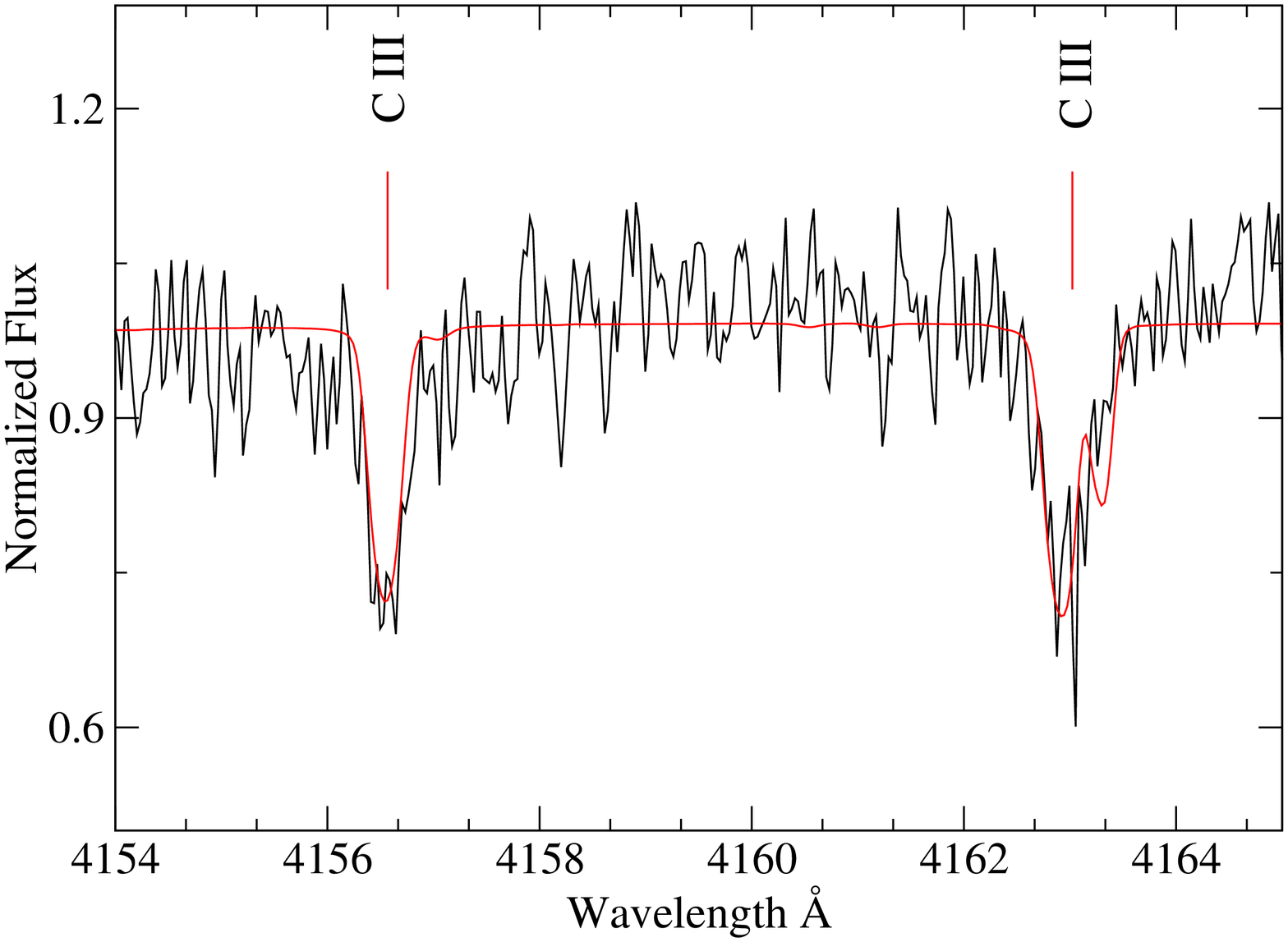}
\includegraphics[trim=2cm 0cm 0cm 0cm, clip=true,angle=-90,scale=0.3]{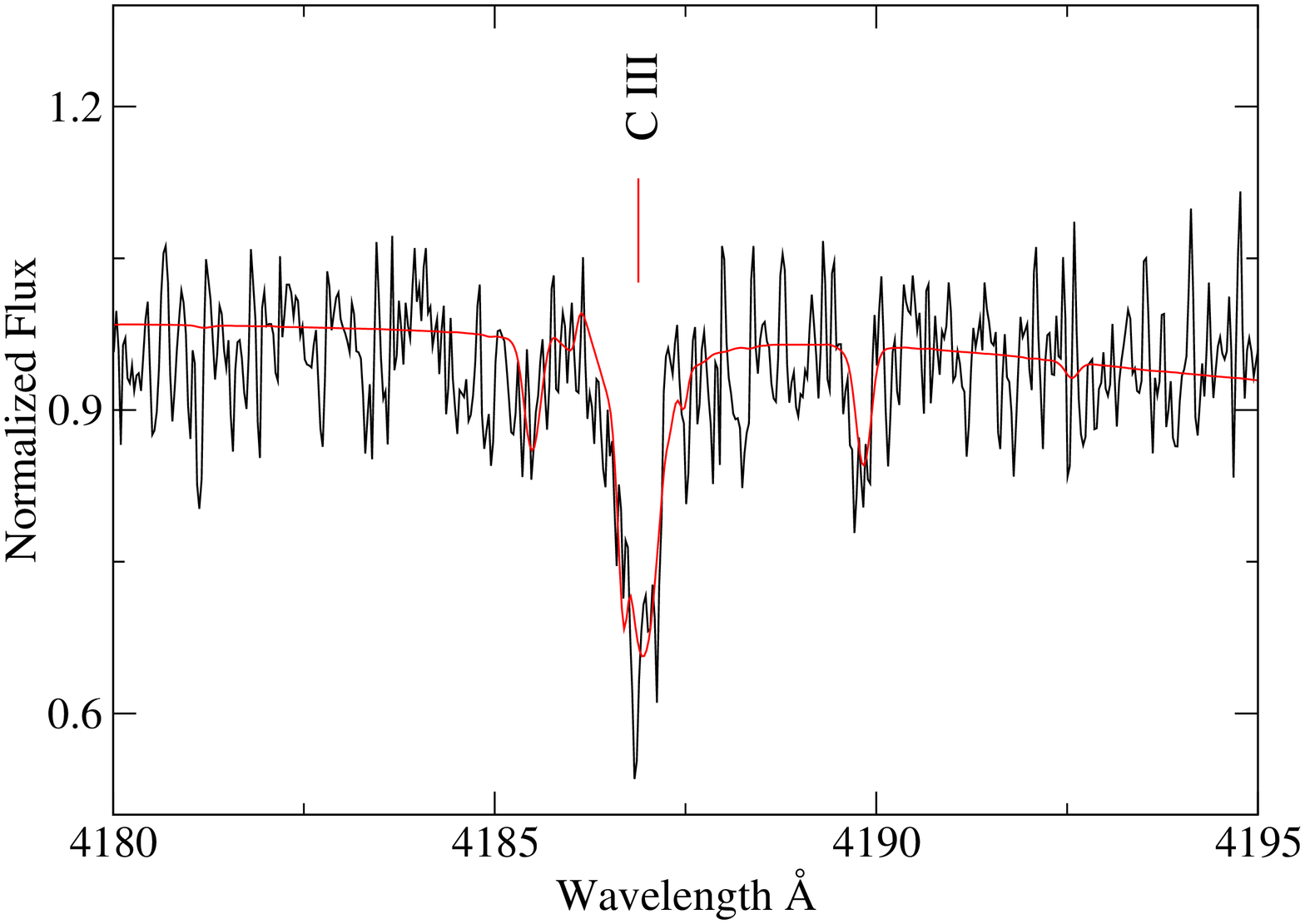}
\includegraphics[trim=2cm 0cm 0cm 0cm, clip=true,angle=-90,scale=0.3]{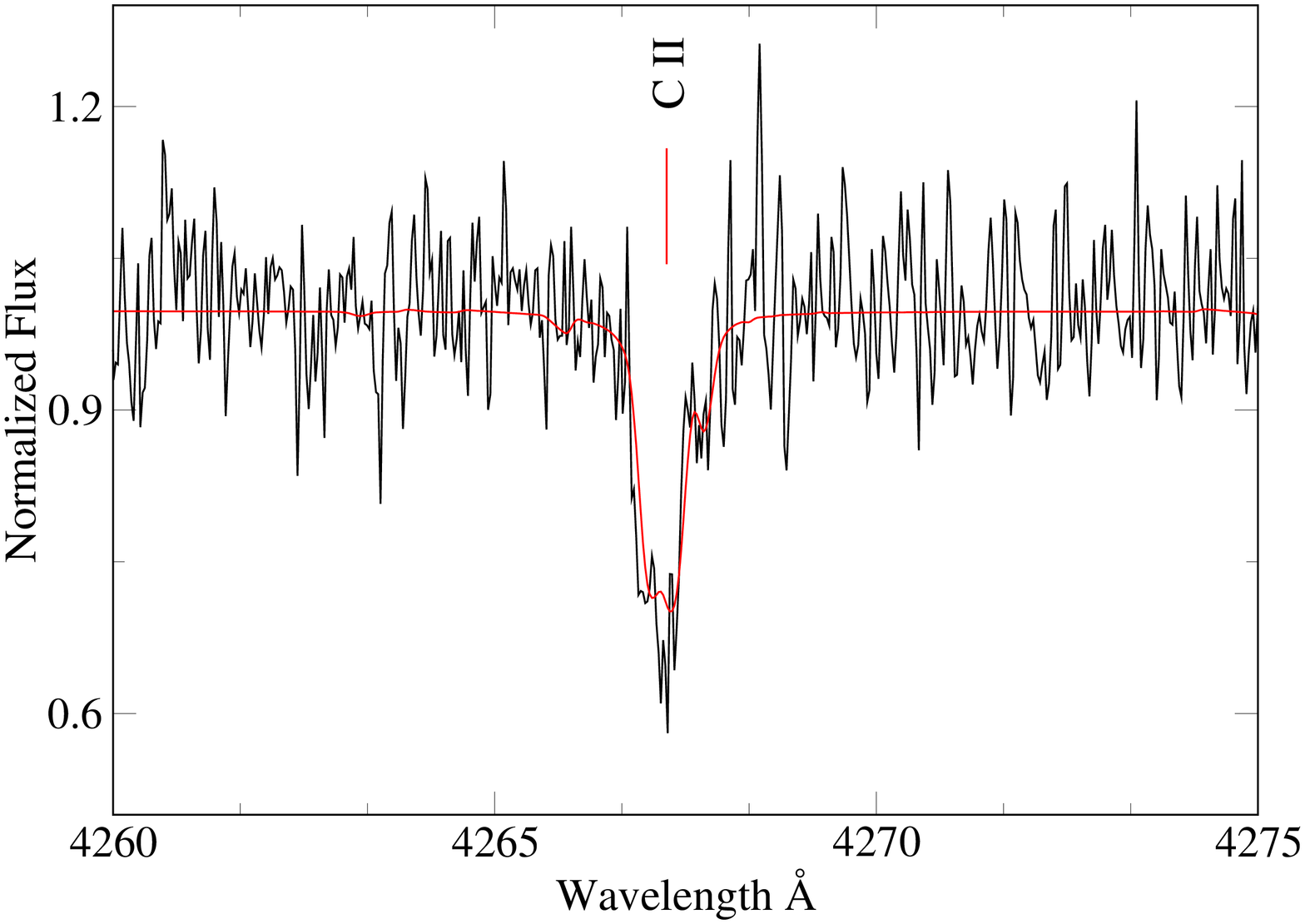}
\includegraphics[trim=2cm 0cm 0cm 0cm, clip=true,angle=-90,scale=0.3]{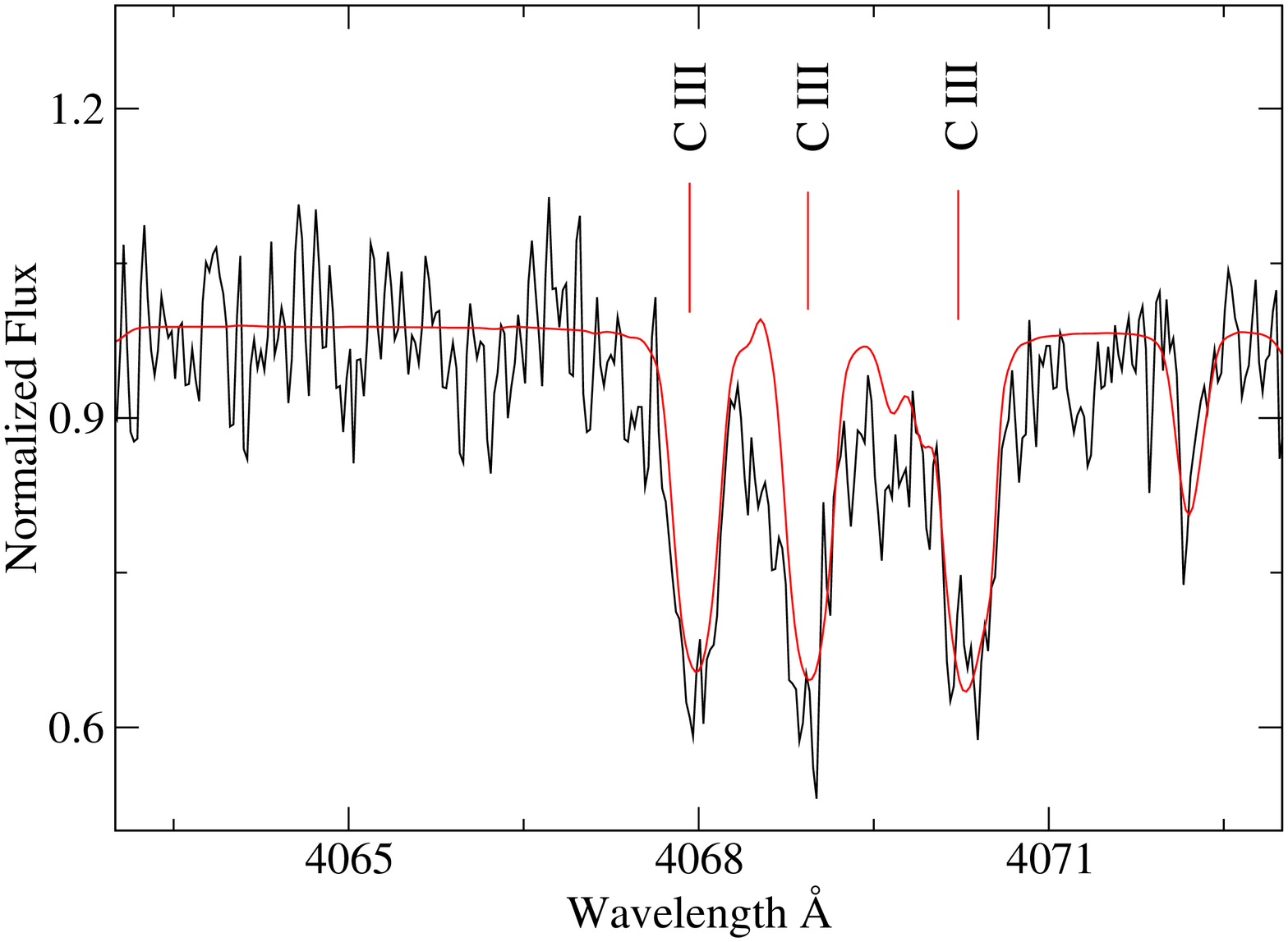}
\includegraphics[trim=2cm 0cm 0cm 0cm, clip=true,angle=-90,scale=0.3]{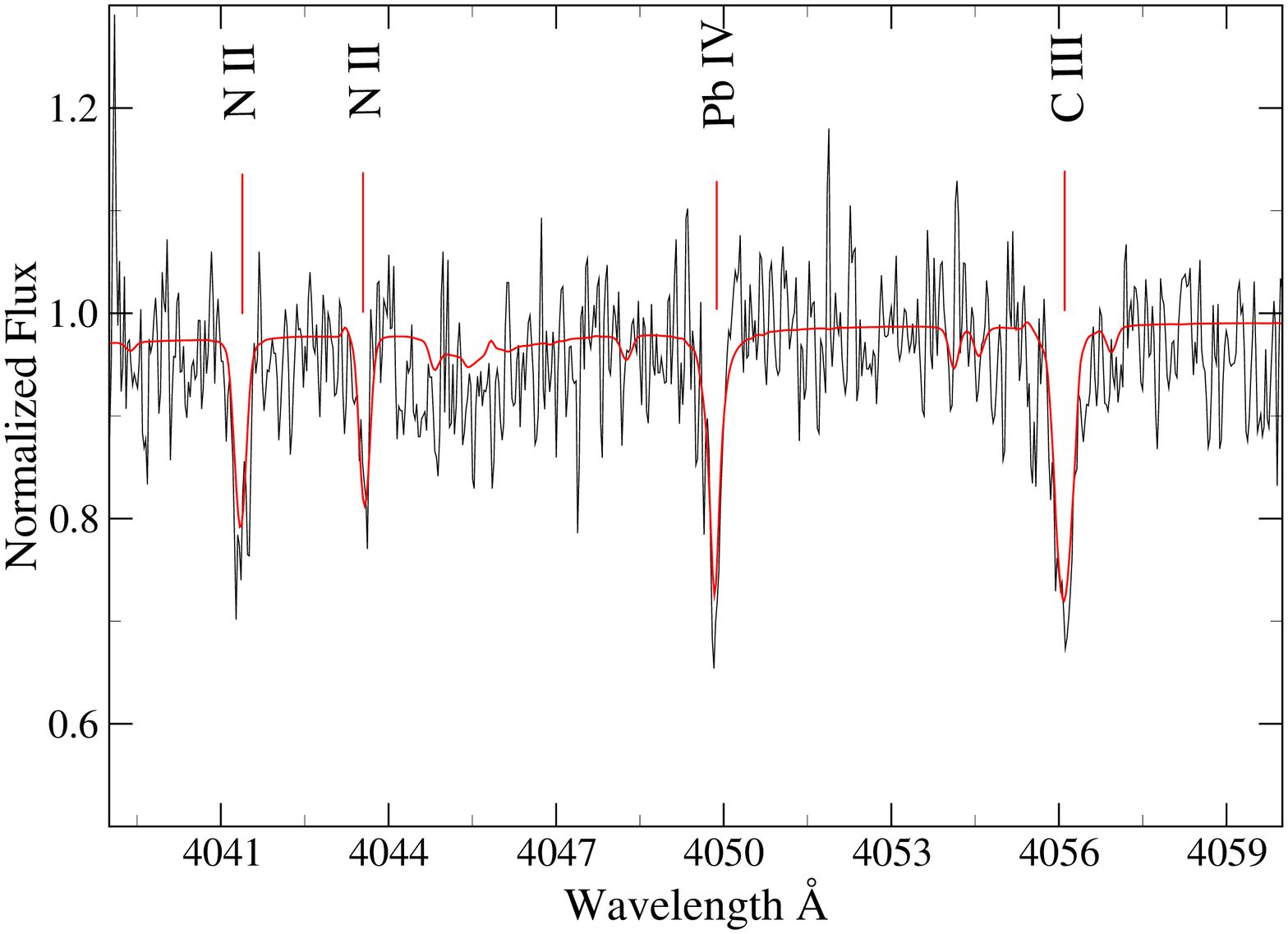}
\includegraphics[trim=2cm 0cm 0cm 0cm, clip=true,angle=-90,scale=0.3]{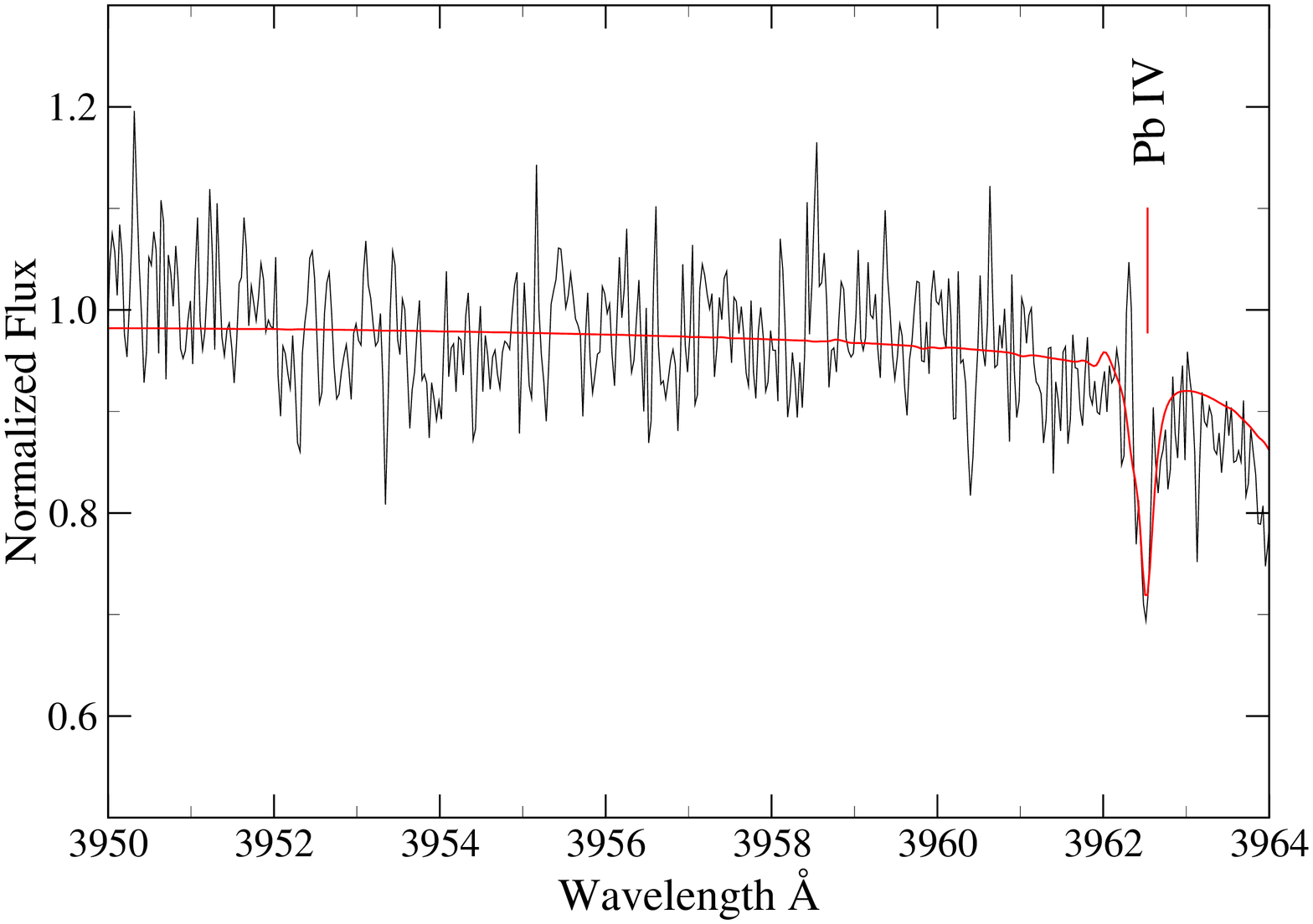}
 \caption{Segments of the VLT UVES spectrum of HE\,1256--2738, 
together with the best-fit model, showing C{\sc iii},  N{\sc ii} and  Pb{\sc iv} lines. 
}
  \label{f:1256}
\end{figure*}

\begin{figure*}
\epsfig{file=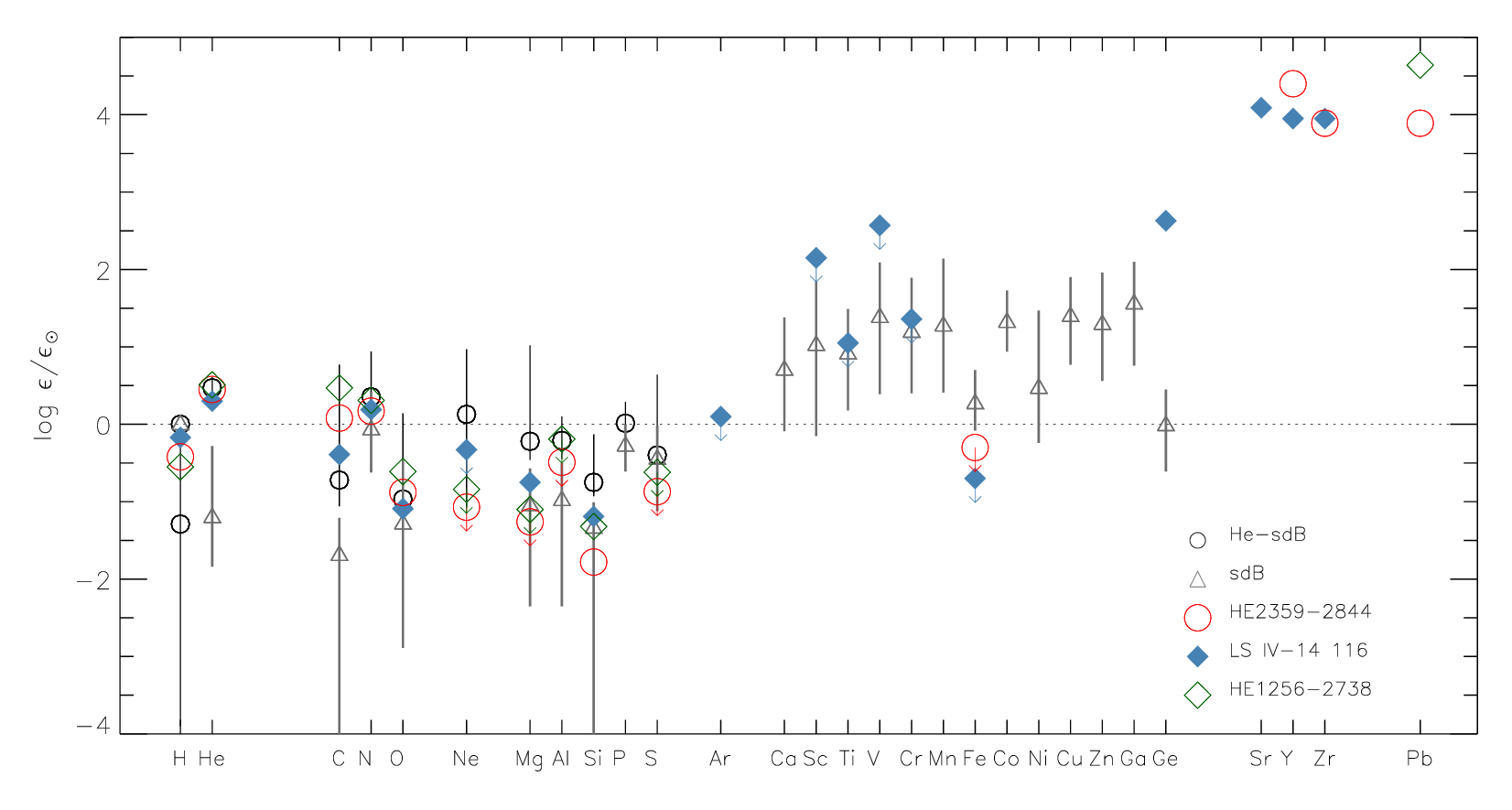, angle=0, width=18cm}
  \caption{Elemental abundances for HE\,2359--2844, HE\,1256-2738 and
    LS\,IV$-14^{\circ}116$ relative to solar values. Abundances for
    He-sd's \citep{naslim10, naslim11} and normal sdBs \citep{pereira11}
    are indicated by a mean (symbol) and range (line).}
  \label{abund}  
\end{figure*}

\section{Abundances}

For abundance measurements the grid model atmospheres closest to the measured $T_{\rm eff}$, $\log
g$, and $n_{\rm He}$ and 1/10 solar metallicity were adopted. After measuring the equivalent widths
of all C, N, O, Mg, Al, Si, and S lines using the spectrum analysis tool DIPSO, the individual line
abundances were calculated using the LTE radiative transfer code SPECTRUM \citep{jeffery01b}.

Mean abundances for each element are given in Table~\ref{t_abunds}. Abundances are given in the form
$\epsilon_i = \log n_i + c$ where $\log \Sigma_i a_i n_i = 12.15$ and $a_i$ are atomic weights. This
form conserves values of $\epsilon_i$ for elements whose abundances do not change, even when the
mean atomic mass of the mixture changes substantially. The errors given in Table~\ref{t_abunds} are
based on the standard deviation of the line abundances about the mean or, in the case of a single
representative line, on the estimated error in the equivalent width measurement. The elemental
abundances shown in Table~\ref{t_abunds} are the mean abundances of all individual lines of an ion.
Abundances for two other intermediate He-sds JL\,87 and LS\,IV$-14^{\circ}116$ are also shown.

In HE\,2359--2844, HE\,0111--1526, HE\,2218--2026 and HE\,1256--2738, carbon is nearly solar or
slightly over abundant, while only upper limits to carbon abundances can be measured for
HE\,1310--2733, HE\,1135--1134, HE\,1136--2504, HE\,1238--1745 and HE\,1258+0113. In all nine stars,
silicon appears to be underabundant relative to solar. Where detectable, sulphur and magnesium are
relatively normal. Upper limits were estimated for several elements by assuming
that, where no lines of a given ion could be measured, the equivalent width of the strongest lines
due to that ion between 4000 and 5000 \AA\  were less than 5 m\AA.

The Zr{\sc iv} and Pb{\sc iv} lines in HE\,2359--2844, together with best-fit theoretical spectra,
are shown in Figs.~\ref{f:2359a} and \ref{f:2359b}. 
The C{\sc ii,iii} and Pb{\sc iv} 3962.48, 4049.80 lines in the spectrum of HE\,1256--2738, 
along with the best fit theoretical spectrum are shown in Fig.~\ref{f:1256}. 
Pb{\sc iv} 4496.2 is not seen in HE1256-2738 because of the lower signal-to-noise
of that spectrum. Pb{\sc iv} 4534.6 and 4605.4 lie in a gap between two sections of the UVES
spectrum. 

Individual line abundances for ytttrium, zirconium and lead for both stars are
given in Table~\ref{t_abunds2}. 
These are all nearly 4\,dex above solar. 
A comparison of abundances relative to solar values for both lead-rich stars and
LS\,IV$-14^{\circ}116$ is shown in Fig.~\ref{abund}. This figure also shows the mean abundances and
ranges for "normal" sdB stars \citep{pereira11} and other helium-rich sdB stars
\cite{naslim10,naslim11}.

\begin{figure}
\epsfig{file=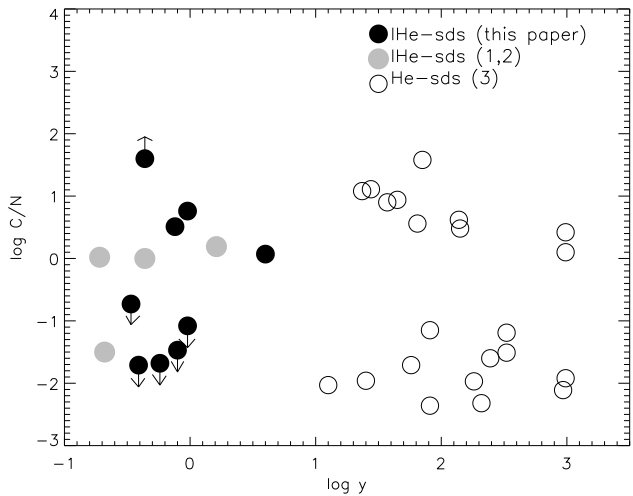, angle=0, width=9cm}
  \caption{Carbon to nitrogen abundance ratio ($n_{\rm C}/n_{\rm N}$)
    as a function of helium  abundance ($y \equiv \log n_{\rm He}/n_{\rm H}$) 
    in helium-rich subdwarfs. This plot includes abundances from
    1. \citet{naslim11, naslim12}, 2. \citet{ahmad07},
    3. \citet{hirsch09.thesis}. Stars where carbon (or nitrogen) is not 
   observed are represented by upper (or lower) limits. }
  \label{logy-logc}  
\end{figure}

\section{Evolutionary status and diffusion in helium-rich hot subdwarfs}

The question posed at the outset was to establish the relationship between the extreme-helium, the
intermediate-helium and the normal hot subdwarfs. With new data from nine additional stars in the
intermediate class we can make some useful observations.

The first task is to establish whether the intermediate and extreme-helium subdwarfs form a single
continuum, or a number of distinct classes. Figure~\ref{logy-logc} shows the carbon-to-nitrogen
ratio as a function of helium-to-hydrogen ratio for all stars analysed to date. It was already
evident that the extreme helium subdwarfs appear to form carbon-rich and carbon-poor groups
\citep{stroeer07,hirsch09.thesis}; this conclusion is further supported by \citep[Fig. 9f]{nemeth12}. In the
double-white dwarf merger model, the distinction can be directly attributed to mass, the carbon-rich
subdwarfs having masses $>0.7 {\rm M_\odot}$ \citep{zhang12a}.

Amongst the intermediate-helium subdwarfs, there is less evidence for two distinct classes in terms
of C/N ratio. The absence of carbon or nitrogen lines in some cases is a problem which needs to be
addressed using higher-quality data. 

\begin{figure*}
\epsfig{file=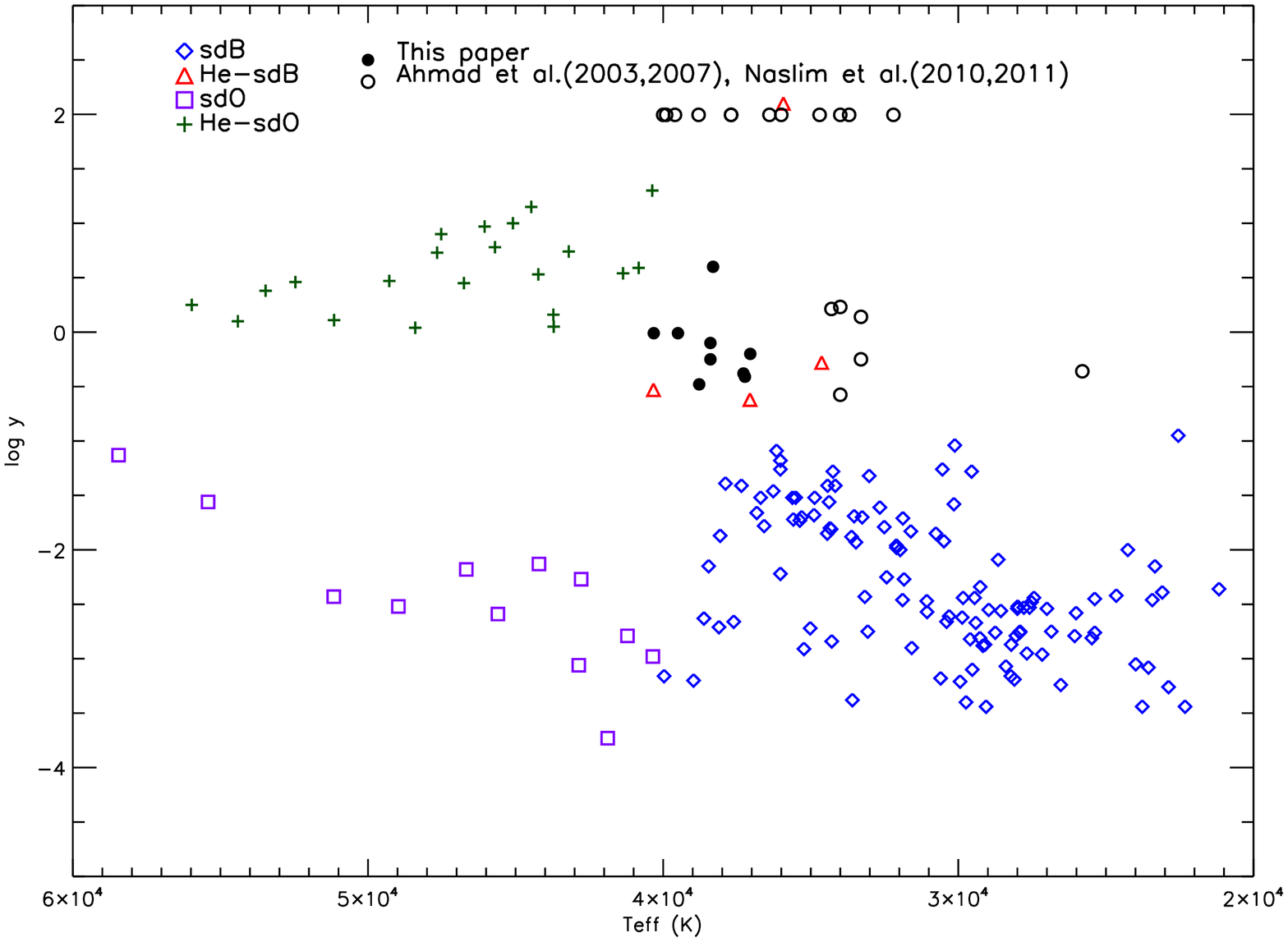, angle=90, width=15cm,trim=0.3cm 0cm 0cm 0cm, clip=true}
  \caption{Helium abundance ($\log y$) as a function of effective temperature
    for subdwarfs analysed by \citet{nemeth12}, and for intermediate helium subdwarfs 
   analyzed by us in the present paper and previously, as indicated. 
   Extreme helium-rich subdwarfs for which the hydrogen abundance is below the 
  detection threshold are shown with $\log y=2$. }
  \label{f:yt}  
\end{figure*}

\citet{nemeth12} describe a group of He-sdO's with $0 < \log y < 1$ and $T_{\rm eff} > 40\,000$\,K.
Most of our stars have $-1 < \log y < 0.5$ and $T_{\rm eff} \leq 40\,000$\,K; most do lie to the
right of 38\,000\,K boundary shown in \citet{nemeth12} (Fig.\,6), but  in the sparsely
populated region between the normal sdBs and He-sdOs. The \citet{nemeth12} sample 
is compared with our own in our Fig.~\ref{f:yt}. Systematic differences 
between the methods of analysis will have some influence;   these should be comparable in 
magnitude with the differences noted in Table 2 between our results and those of \citet{stroeer07}. 
Note how the inclusion of the intermediate helium-rich subdwarfs strengthens parallels drawn between 
the $\log y - T_{\rm eff}$ diagram and the helium class -- spectral type diagram in the hot subdwarf 
classification described by \citet{drilling13}. 

It is evident that some of these {\it intermediate} helium-rich stars are extremely peculiar. As far
as we can tell, HE\,2359--2844 and HE\,1256--2738 are the most lead-rich stars known to science.
LS\,IV$-14^{\circ}116$ shows $4$ dex overabundances in zirconium, strontium, and yttrium (and 3 dex
in germanium) \citep{naslim11}. Not having reached the thermally-pulsing asymptotic giant branch, a
nuclear origin for the excess in these s-process elements appears unlikely. This leaves
radiatively-driven diffusion as a favoured explanation since it is theoretically capable of
concentrating particular species into a thin line-forming layer of the photosphere.

\citet{geier13} reports metal abundances for a large sample of normal subdwarf B stars. In general
and with some dependence upon effective temperature, most elements heavier than helium and lighter
than calcium, but excepting nitrogen, are depleted by a factor $\approx 10$ relative to solar.
Elements heavier than and including calcium are enhanced by a similar amount, excepting iron, which
is solar, and vanadium, which is $>3$ dex enhanced. The intermediate helium-rich sdB stars
UVO\,0512--08 and PG\,0909+276, analysed by \citet{edelmann03th}, show $>3$ dex overabundances in
scandium, titanium, vanadium, manganese and nickel. Other elements (Ga, Sn, Pb) have been measured as $>2$ dex overabundant in normal
sdB stars \citep{otoole06} using ultraviolet spectroscopy.

Calculations of the evolution of horizontal-branch stars including radiatively-driven diffusion by
\citet{michaud11} reproduce moderately well the surface abundances of normal sdB stars (their Fig.
5), at least for elements included in the opacities \citep{iglesias96}. Diffusion calculations for
most elements observed with excessive overabundances ({\it e.g.} Ge, Sr, Y, Zr, Sn and Pb) are not
yet possible. The \citet{michaud11} calculations, however, represent equilibrium abundances for
stars established on the (extreme) horizontal branch (EHB). Most intermediate helium-rich sdB stars 
lie above the classical EHB, possibly because they are evolving onto {\it or} away from a stable core-helium-burning
configuration. In either case, the equilibrium result may not be valid. For example, \citet{hu11}
computes a time-dependent model which demonstrates that it takes 3000 years or so after radiative
levitation commences for the surface layer chemistry to stabilise, whilst \citet{groth85} suggest
that increased photospheric mixing and helium enrichment occurs as a hot subdwarf evolves away from
the extended horizontal branch.

It could be argued that stars having excessive overabundances of selected elements provide evidence
for stratification immediately following onset of conditions which allow radiative levitation to operate, {\it
i.e.} on approach to the EHB. In the absence of detailed calculations, it could also be
argued that, as a star evolves away from the EHB, photospheric conditions may change sufficiently
for established concentrations to rise (or sink) into the line-forming region. 

In this context, one
might ask whether there is some ``sweet spot'' in effective temperature and surface gravity where
particular elemental overabundances are likely? So far, the evidence is weak; LS IV$-14^{\circ}116$
is 3\,000\,K cooler than HE\,2359--2844 but shows a similar zirconium abundance, whilst the two
lead-rich stars also differ in effective temperature by 3\,000\,K. Consequently, there must be
additional factors, such as age or mass, which lead to the formation of a ``heavy metal'' star.

\section{Conclusion}

We have analysed high-resolution spectra of nine intermediate helium-rich hot subdwarfs, 
having helium-to-hydrogen ratios between 0.5 and 5 (by number). 
In terms of carbon and nitrogen abundances and unlike the more extreme helium-rich subdwarfs, 
we do not find evidence of separate carbon- and nitrogen-rich groups. 

Two stars, HE\,2359--2844  and  HE\,1256--2738, show 
absorption lines due to triply ionized lead (Pb{\sc iv}) which have never previously been detected 
in any star.  From these lines, we have measured an atmospheric 
abundance of lead  which is nearly  ten thousand times that measured in the Sun. To our knowledge, 
these are the most lead-rich stars known to science. 

The lead abundance is also ten to 100 times that previously measured in normal hot subdwarf
atmospheres from ultraviolet spectroscopy. HE\,2359--2844 also shows zirconium and yttrium
abundances similar to those in the zirconium star LS IV$-14^{\circ}116$. The best physical
explanation for the large overabundances in these stars is that selective radiative forces levitate
and concentrate
specific ions into thin layers in the photosphere; these layers should coincide with regions where
the specific opacity of an ion has a maximum value. Where high concentrations coincide with the
line-forming layer, overabundances will be observable.

Extreme overabundances are only seen in intermediate helium-rich subdwarfs, which are themselves
intrinsically rare and overluminous compared with normal subdwarfs. The latter are stable
helium-core-burning stars with helium-poor surfaces. This suggests an association between the
process that transforms a helium-poor subdwarf into an intermediate helium-rich subdwarf (or vice
versa) {\it and} the process that produces excessive overabundances of exotic elements. We suggest
that organized stratification of the atmosphere is more likely to occur during transition from a
new-born helium-rich subdwarf to a normal helium-poor subdwarf.

\section*{Acknowledgments}
The Armagh Observatory is funded by a grant from the Northern Ireland 
Dept of Culture, Arts and Leisure.

\bibliographystyle{mn2e}
\bibliography{ehe}

\appendix
\section{Atomic data}
In our previous paper \citep{naslim11}, we discussed several lines of Zr{\sc iv}, including
\begin{center}
 4198.26 \AA: 5d $^2$D$_{5/2}$ -- 6p $^2$P$^{\rm o}_{3/2}$ \\
 4317.08 \AA: 5d $^2$D$_{3/2}$ -- 6p $^2$P$^{\rm o}_{1/2}$ .
\end{center}
We now need atomic data for two further lines:
\begin{center}
 3686.90 \AA: 6p $^2$P$^{\rm o}_{3/2}$ -- 6d $^2$D$_{5/2}$ \\
 3764.31 \AA: 6d $^2$D$_{5/2}$ -- 6f $^2$F$^{\rm o}_{7/2}$
\end{center}
For the first of these new transitions, we were able to make use of
the wave functions obtained in calculating data for the previously
studied transitions, since the 6p levels were again included and the
6d orbital function was in fact optimised on the energy of the 6d $^2$D
state, even though it was introduced to provide some valence shell
correlation for the 5d -- 6p transitions.  Here we augmented the
4s$^2$4p$^6$6d configuration by the configurations
4s$^2$4p$^2$4d$^2$6d and 4s4p$^6$4d6d, in parallel with the
configurations introduced for the 4d and 5d states. \\[0.2cm] The
second of the new transitions required the generation of 4f, 5f, 6f
orbitals.  We found that if we generated the orbitals using only these
three configurations, there was significant mixing between them when
correction configurations were added.  So in our optimisation, we
incorporated the main correlation configurations as well as the
dominant configurations which are of the form 4s$^2$4p$^6$nf.
Additionally, we optimised 7f on the 4s$^2$4p$^6$7f $^2$F$^{\rm o}$
state, to allow for further valence shell correlation. \\[0.2cm] In
the final calculations, we used the following sets of configurations:
\\[0.2cm]
\begin{tabular}{lcl}
Even parity & : & 4s$^2$4p$^6$nd, (4 $\leq$ n $\leq$7) \\
            &   & 4s$^2$4p$^4$4d$^2$nd, (4 $\leq$ n $\leq$7) \\
            &   & 4s4p$^6$(4dnd + kpmp), \\
            &   & (4 $\leq$ n $\leq$7, 5 $\leq$ k,m $\leq$ 8) \\
            &   & 4s$^2$4p$^5$(4d5p+4d6p+4dnf), (4 $\leq$ n $\leq$7) \\[0.2cm]
Odd parity  & : & 4s$^2$4p$^6$(5p+6p+nf), (4 $\leq$ n $\leq$7) \\
            &   & 4s$^2$4p$^4$4d$^2$(5p+6p+nf), (4 $\leq$ n $\leq$7) \\
            &   & 4s4p$^6$(4dmp+4dnf + kp$l$d),  \\
            &   & (m=4,5; 4 $\leq$ n $\leq$7; 5 $\leq$ k $\leq$ 8); $l$=5,6) \\
            &   & 4s$^2$4p$^5$4dnd, (4 $\leq$ n $\leq$7) \\[0.2cm]
\end{tabular}
\newline
\noindent
The calculated atomic data for the two key transitions are as follows: \\[0.2cm]
\begin{tabular}{ccll}
Wavelength (\AA) & Transition & $f_l$ & $gf$ \\ \hline \\
 3686.90 & 6p $^2$P$^{\rm o}_{3/2}$ -- 6d $^2$D$_{5/2}$ & 1.3938 & 5.575 \\
 3764.31 & 6d $^2$D$_{5/2}$ -- 6f $^2$F$^{\rm o}_{7/2}$ & 0.5098 & 3.059 \\[0.2cm]
\end{tabular}
\newline
\noindent
where $f_l$ is the oscillator strength, calculated in length form, and
the $g$ of $gf$ is the (2J+1) value of the lower level of the
transition.  These data are calculated with experimental energies, as
given by NIST.  The calculated transition energies are within 2\% of
the experimental values.  In view of this, and of the close agreement
between the length and velocity forms of the oscillator strengths, we
anticipate that these data should be correct to within about 10\%.

For this paper, we additionally required oscillator strengths for a number of 
lines of Pb{\sc iv} not previously observed in any star. Recent 
calculations of these quantities have been carried out by \citet{safranova04} 
and \citet{alonso11}. The former carried out a Many-Body Perturbation Theory (MBPT)
calculation,  based on single and double excitations of Dirac-Fock  orbitals, 
whilst the latter give results obtained using the relativistic Hartree-Fock method
\citep{cowan81}. The two sets of results differ by about a factor of two.  We have 
chosen to use the third-order MBPT values for several reasons: first, the MBPT
method has proved in general to be of greater accuracy; second, if we take as
an example the line at $\lambda$3963.5,   \citet{safranova04}  attribute this to a
one-electron 6d--7p transition, whereas  \citet{alonso11} attribute it to a two-electron
5d$^{10}$6d -- 5d$^9$6s6p transition; since the dipole operator involved in evaluating
$f$-values is a one-electron operator, substantial configuration mixing would be required
for this latter assignment, and this doesn't seem likely. 

\label{lastpage}

\end{document}